\documentclass[%
 reprint,
 nobibnotes,
 amsmath,amssymb,
 aps,
 prb,
]{revtex4-1}

\usepackage{siunitx}
\usepackage{bm}%
\usepackage{xcolor} 
\usepackage{graphicx}
\usepackage[utf8]{inputenc}
\usepackage{epstopdf}
\usepackage{afterpage}
\usepackage{setspace}
\usepackage{booktabs}
\usepackage[colorlinks=true,linkcolor=blue,citecolor=blue]{hyperref}
\usepackage{subcaption}
\usepackage{float}
\usepackage{multirow}

\usepackage{mhchem}            
\usepackage[utf8]{inputenc}     
\usepackage[T1]{fontenc}        
\DeclareUnicodeCharacter{0308}{\"}  

\usepackage{natbib}

\expandafter\ifx\csname package@font\endcsname\relax\else
 \expandafter\expandafter
 \expandafter\usepackage
 \expandafter\expandafter
 \expandafter{\csname package@font\endcsname}%
\fi
\begin{document}

\setstretch{1.0}
\title{A High-Dimensional Neural Network Potential for Co$_3$O$_4$}

\author{Amir Omranpour}
\affiliation{Lehrstuhl f\"ur Theoretische Chemie II, Ruhr-Universit\"at Bochum, 44780 Bochum, Germany}
\affiliation{Research Center Chemical Sciences and Sustainability, Research Alliance Ruhr, 44780 Bochum, Germany}
\author{J\"org Behler}
\email{joerg.behler@rub.de}
\affiliation{Lehrstuhl f\"ur Theoretische Chemie II, Ruhr-Universit\"at Bochum, 44780 Bochum, Germany}
\affiliation{Research Center Chemical Sciences and Sustainability, Research Alliance Ruhr, 44780 Bochum, Germany}
\date{\today}

\begin{abstract}

The Co$_3$O$_4$ spinel is an important material in oxidation catalysis. Its properties under catalytic conditions, i.e., at finite temperatures, can be studied by molecular dynamics simulations, which critically depend on an accurate description of the atomic interactions. Due to the high complexity of Co$_3$O$_4$, which is related to the presence of multiple oxidation states of the cobalt ions, to date \textit{ab initio} methods have been essentially the only way to reliably capture the underlying potential energy surface, while more efficient atomistic potentials are very challenging to construct. Consequently, the accessible length and time scales of computer simulations of systems containing Co$_3$O$_4$ are still severely limited. Rapid advances in the development of modern machine learning potentials (MLPs) trained on electronic structure data now make it possible to bridge this gap. In this work, we employ a high-dimensional neural network potential (HDNNP) to construct a MLP for bulk Co$_3$O$_4$ spinel based on density functional theory calculations. After a careful validation of the potential, we compute various structural, vibrational, and dynamical properties of the Co$_3$O$_4$ spinel with a particular focus on its temperature-dependent behavior, including the thermal expansion coefficient. 
\end{abstract}

\maketitle
\section{Introduction}\label{sec:introduction}

Cobalt oxide (Co$_3$O$_4$) is a transition metal oxide with mixed valence states that has gained considerable attention in recent years due to its distinct chemical and physical properties, which hold significant promise for a wide range of technological applications. These include its use as a catalyst in processes like alcohol oxidation \cite{falk2021identification}, water oxidation\cite{jiao2009nanostructured}, methane combustion\cite{hu2008selective}, and CO oxidation\cite{xie2009low}, but also as a component in lithium-ion batteries and gas sensors\cite{li2005co3o4}. Among its various forms, in particular, the Co$_3$O$_4$ spinel structure has been extensively studied for its ability to promote oxidation reactions, especially in the selective oxidation of hydrocarbons\cite{waidhas2020secondary, hill2017site, finocchio1997ftir}. Additionally, its electronic, magnetic, and redox properties have motivated numerous studies in recent years\cite{omranpoorLowIndex,anke2019selective, doheim2002catalytic, yang2021selective, omranpoor20232, anke2020reversible, nano12060921, douma_Probing_2022,omranpoor2022influence}. Despite these efforts, the atomistic mechanisms of chemical processes involving Co$_3$O$_4$ in catalysis and electrochemistry have not yet been fully understood. Hence, further research employing realistic model systems is needed to gain a deeper atomistic understanding of Co$_3$O$_4$ to fully explore and expand its potential applications.

\begin{figure}[ht!]
    \centering
    \begin{subfigure}[b]{0.42\textwidth}
        \centering
        \includegraphics[width=\textwidth,trim= 250 150 250 200,clip=true]{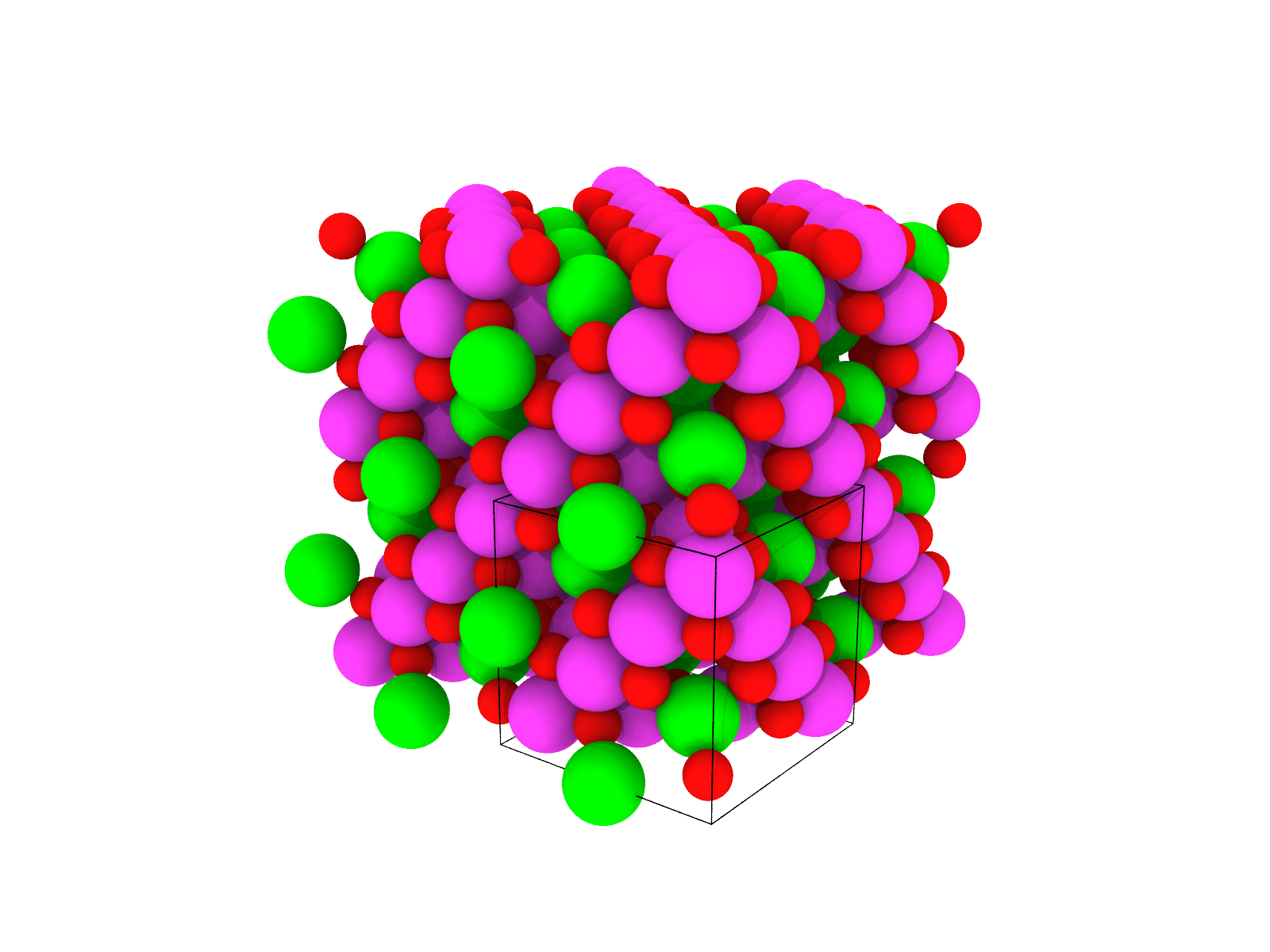}
        \caption[]%
        {{Co$_3$O$_4$ (2$\times$2$\times$2) Supercell}}    
        \label{fig:unit_cell}
    \end{subfigure}
    \begin{subfigure}[b]{0.1\textwidth}  
        \centering 
        \includegraphics[width=1\textwidth,trim= 280 0 370 0,clip=true]{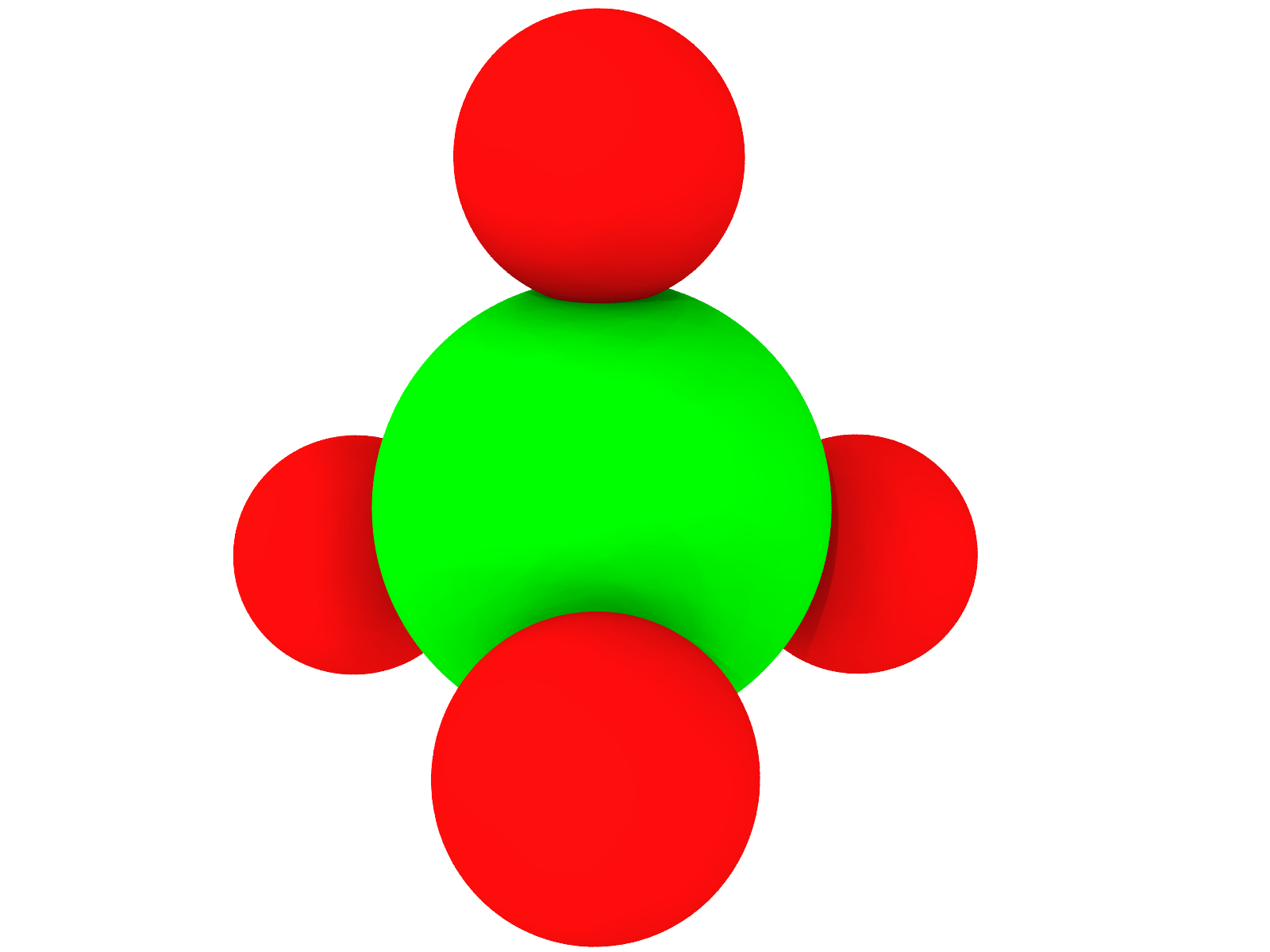}
        \caption[]%
        {{Co\textsuperscript{2+}}}    
        \label{fig:co2+}
    \end{subfigure}
    \hfill
    \begin{subfigure}[b]{0.1\textwidth}   
        \centering 
        \includegraphics[width=1.2\textwidth,trim= 200 0 260 0,clip=true]{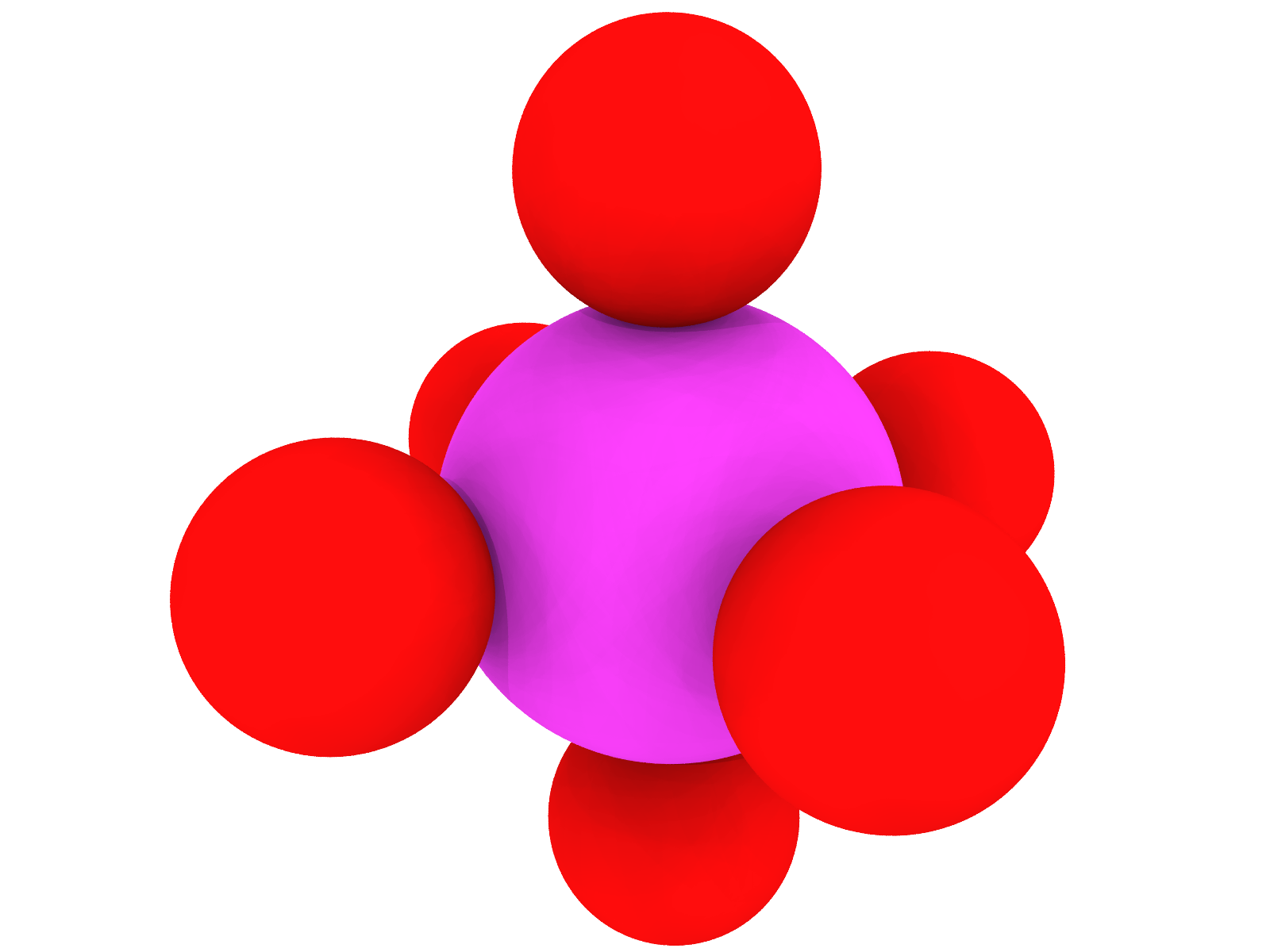}
        \caption[]%
        {{Co\textsuperscript{3+}}}    
        \label{fig:co3+}
    \end{subfigure}
    \hfill
    \begin{subfigure}[b]{0.1\textwidth}   
        \centering 
        \includegraphics[width=1.1\textwidth,trim= 280 0 310 0,clip=true]{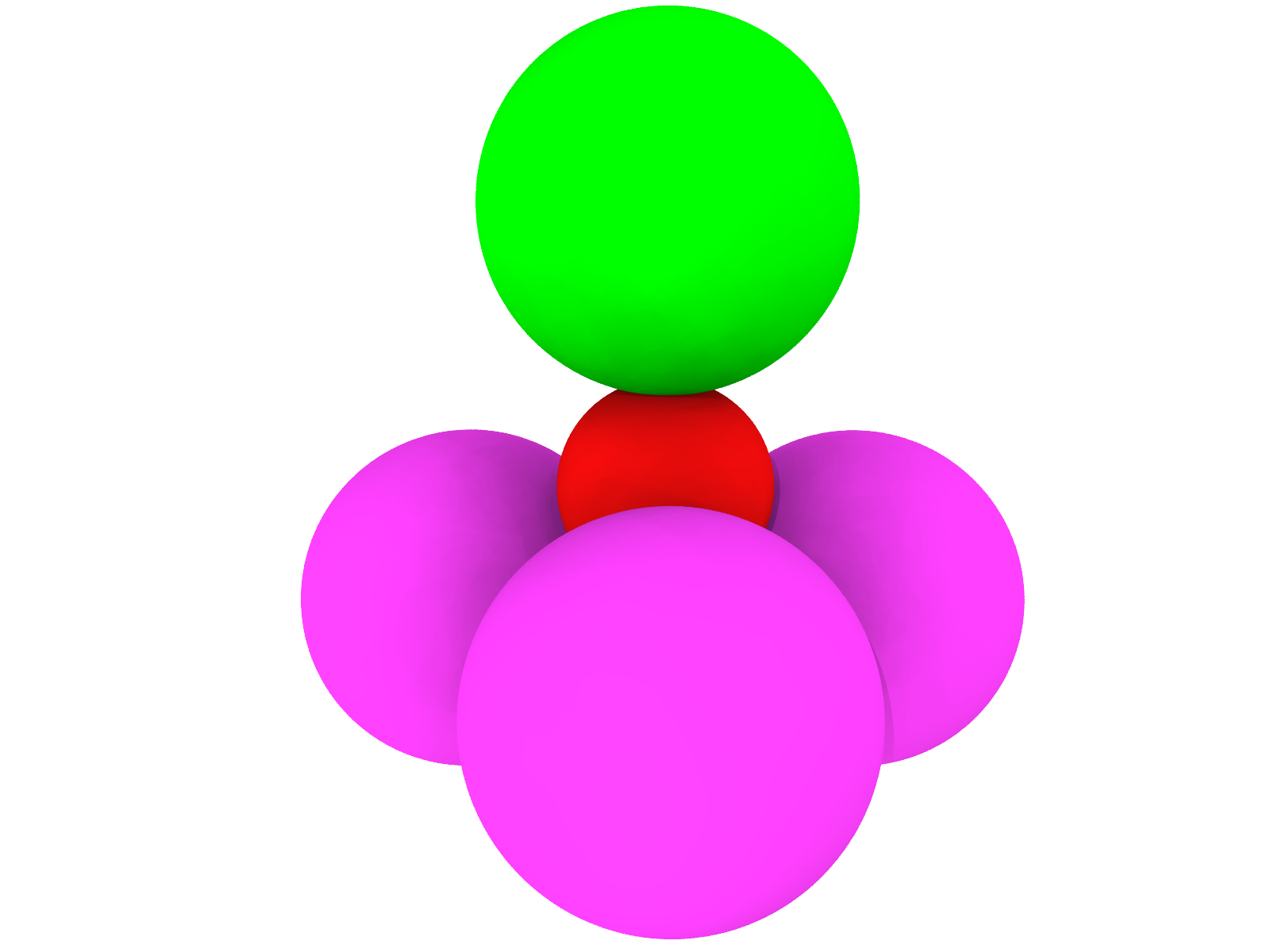}
        \caption[]%
        {{O\textsuperscript{2-}}}    
        \label{fig:o2-}
    \end{subfigure}
    \caption[]{Structure of a spinel Co$_3$O$_4$ (2$\times$2$\times$2) supercell with the unit cell denoted by the black box, (a) consisting of tetrahedrally coordinated Co\textsuperscript{2+} ions (b), octahedrally coordinated Co\textsuperscript{3+} ions (c), and tetrahedrally coordinated O\textsuperscript{2-} ions (d) (Co\textsuperscript{2+} (green), Co\textsuperscript{3+} (magenta), and O\textsuperscript{2-} (red)).}
    \label{fig:Co3O4_structure}
\end{figure}

Structurally, Co$_3$O$_4$ crystallizes in a cubic normal spinel configuration, where cobalt ions are found in two different oxidation states, Co\textsuperscript{2+} and Co\textsuperscript{3+}. These ions occupy interstitial tetrahedral and octahedral sites, respectively, within a face-centered cubic (FCC) lattice formed by the oxygen ions (Figure~\ref{fig:Co3O4_structure}). This dual valency is responsible for its complex behavior and contributes to its redox activity, making Co$_3$O$_4$ a highly interesting material for applications in catalysis. For simplicity, and as is common in the literature on Co$_3$O$_4$, we adopt the terminology of using Co\textsuperscript{2+} and Co\textsuperscript{3+} interchangeably with tetrahedrally coordinated and octahedrally coordinated Co ions, respectively, unless otherwise noted. Although this naming convention is not entirely accurate, it is convenient for this work, which is concerned solely with the bulk Co$_3$O$_4$ spinel. 

\begin{figure}[hb!]
        \centering
        \begin{subfigure}[b]{0.3\textwidth}
            \centering
            \includegraphics[width=\textwidth,trim= 0 0 0 0,clip=true]{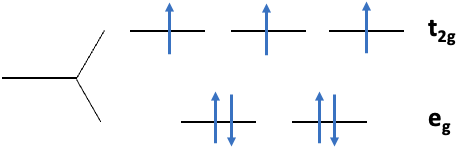}
            \caption[]%
            {{Co\textsuperscript{2+} $d^7$ configuration}}    
            \label{fig:mean and std of net14}
        \end{subfigure}
        \hfill
        \begin{subfigure}[b]{0.3\textwidth}  
            \centering 
            \includegraphics[width=\textwidth,trim= 0 0 0 0,clip=true]{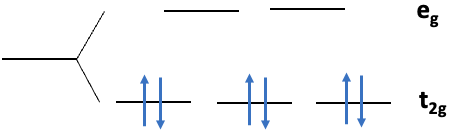}
            \caption[]%
            {{Co\textsuperscript{3+} $d^6$ configuration}}    
            \label{fig:mean and std of net24}
        \end{subfigure}
	\caption[]{Schematic representation of the crystal field splitting of tetrahedrally coordinated Co\textsuperscript{2+} ions in the $d^7$ configuration (a) and octahedrally coordinated Co\textsuperscript{3+} ions in the $d^6$ configuration in a low-spin state (b).}\label{fig:CF-split}
\end{figure}

The crystal fields cause the five degenerate atomic $d$ orbitals of cobalt to split into two distinct groups. This results in three unpaired $d$ electrons of Co\textsuperscript{2+} with the experimental magnetic moment of 3.26 $\mu_B$\cite{roth1964magnetic}, whereas Co\textsuperscript{3+} typically adopts a low-spin state configuration with a quenched magnetic moment (Figure~\ref{fig:CF-split}). At room temperature, Co$_3$O$_4$ behaves as a paramagnetic semiconductor, transitioning to an antiferromagnetic state below 40 K\cite{roth1964magnetic}. The antiferromagnetism is mainly due to weak interactions between neighboring Co\textsuperscript{2+} ions. Co$_3$O$_4$ is an intrinsic p-type semiconductor. In the 220–400 K temperature range, conduction primarily occurs through polaronic hopping of holes, while between 170–220 K, it is due to variable-range hopping of holes\cite{cheng1998electrical,koumoto1981electrical}. The Co$_3$O$_4$ band gap is estimated to be around 1.6 eV\cite{kim2003optical, shinde2006supercapacitive}. Although the electronic structure of Co$_3$O$_4$ spinel has been addressed by theoretical studies in detail, most of these studies have been restricted to properties at $T$ = 0 K and thus do not account for dynamic and temperature-dependent properties, which are crucial for chemical processes taking place at higher temperatures, e.g., in catalysis.

To study the finite-temperature properties of materials, e.g. by molecular dynamics simulations, there have traditionally been two major options to compute the required energies and forces, \textit{ab initio} electronic structure methods and approximate empirical potentials. The disadvantage of the former is that \textit{ab initio} methods such as density functional theory (DFT) offer very limited accessible time and length scales -- typically a few hundred atoms and 10-100 ps simulation time for most practical purposes -- due to the high computational costs. On the other hand, empirical potentials, which provide a direct functional relation between the atomic positions and the potential energy of the system, are computationally several orders of magnitude less demanding. However, they suffer from limited accuracy due to the underlying physical approximations typically used in their construction. More importantly, such potentials are available for only a limited number of materials, and for more complex materials, such as Co$_3$O$_4$ that include ions of the same element in multiple oxidation states interacting differently, are very difficult to construct.

In recent years, a new class of interatomic potentials making use of machine learning algorithms has been developed to bridge this gap~\cite{P4885,P5673,P6102,P6112}. Such machine learning potentials (MLP) provide an accuracy close to that of \textit{ab initio} methods while accelerating the speed of simulations by several orders of magnitude, thereby making accessible length and time scales that were previously unattainable. In this work, we present an interatomic potential for Co$_3$O$_4$ employing high-dimensional neural network potentials (HDNNPs)\cite{behler2007generalized,behler2021four}, which are a frequently used class of MLPs, to assess the ability of HDNNPs to describe bulk materials with a complex electronic structure.

To date, the only work addressing the development of an MLP for cobalt oxides has been reported by Kong et al. \cite{kong2019stability} and focuses on the stability and phase transitions of CoO$_x$ phases by reconstructing the global potential energy surface of CoO$_x$ using machine learning. The present work differs from the aforementioned study in two distinct ways: first, the previous work does not focus on Co$_3$O$_4$, but rather on different phases of CoO$_x$ and is primarily concerned with the structure-energy relations of these phases. Moreover, finite-temperature properties or the related dynamics have not been addressed. Another related oxide material containing a transition metal in different oxidation states, Mn$^{3+}$ and Mn$^{4+}$, is LiMn$_2$O$_4$, which has been studied by Eckhoff et al. using HDNNPs~\cite{P5866,P5867,P6141}.

The structure of our work is as follows: first, a brief introduction to the main methodology used in this study is presented. We then continue with the computational details of the techniques used in this work, specifically covering the electronic structure calculations performed and the structure of the datasets used to construct the HDNNP. The details of the neural network settings for constructing the HDNNP are also provided. Finally, the atomistic simulations are described. In the results and discussion section, we discuss the choice of the DFT functional used in this work by demonstrating its ability to reproduce important experimental observables with good accuracy. We then validate the HDNNP by showing that it can accurately reproduce the results of its underlying electronic structure calculations. This is followed by the calculation of several other properties divided into three main types: static calculations, e.g., lattice parameters and elastic constants at $T$ = 0 K, phonon properties such as the phonon band structure and density of states, and finite-temperature properties like the variation of lattice parameters as a function of temperature.
\section{Methods}\label{sec:methods}

Performing atomistic simulations based on machine learning potentials involves three main steps: (1) the reference electronic structure calculations, i.e., DFT in this work; (2) the machine learning potential construction, i.e., the HDNNP training; and (3) molecular dynamics (MD) simulations based on energies, forces and possibly the stress tensor provided by the MLP.

In contrast to \textit{ab initio} MD, where the energy and forces are computed on-the-fly using DFT at each step, MLPs like HDNNPs provide a continuous function representing the potential energy surface of the system that maps the system's total energy to its corresponding atomic configuration. In order to represent the atomic environments, descriptors are utilized as structural fingerprints. The relationship between structure and energy thus relies on these descriptors, which preserve the potential energy surface with respect to translational, rotational, and permutational invariances. In HDNNPs, often atom-centered symmetry functions (ACSF) \cite{behler2011atom} are used as descriptors to ensure that all equivalent representations of a given structure produce the same potential energy.

The chemical environment of an atom, encompassing all neighboring atoms within a cutoff sphere of radius $R_{\mathrm{c}}$, is characterized by a vector of ACSF values. In second-generation HDNNPs~\cite{behler2007generalized}, which do not contain long-range electrostatic interactions, the cutoff radius $R_{\mathrm{c}}$ must be sufficiently large to capture all significant energetic interactions, with typical values ranging from 5 to 10~\AA{}. In this study, two types of ACSFs are utilized, radial and angular symmetry functions, as defined in Ref. \citenum{behler2011atom}. Depending on the system's complexity, between 30 and 150 ACSFs are typically used per atom.

Apart from the chemical elements of the atoms, no additional information such as atom types, fixed oxidation states, or predefined bonds is specified. This makes HDNNPs reactive, enabling them to accurately describe the formation and dissociation of bonds and also the changes of oxidation states in accordance with the underlying electronic structure method. Since the dimension of the ACSF vectors is defined by the chosen symmetry functions and is independent of the particular chemical environment, these vectors can be used as inputs for neural networks requiring a fixed number of input neurons.

A separate atomic neural network is then constructed for each chemical element \(\alpha\), which processes the structural information of the geometric environment of each atom \(n\), resulting in its atomic energy contribution \(E_n^\alpha\). The total potential energy \(E\) of the system, composed of \(N_\text{elements}\) elements and \(N_\text{atoms}^\alpha\) atoms of element \(\alpha\), is then obtained by summing all the atomic energy contributions,

\begin{equation}
E = \sum_{\alpha=1}^{N_\text{elements}} \sum_{n=1}^{N_\text{atoms}^\alpha} E_n^\alpha.
\end{equation}

The atomic energy contributions are determined using atomic neural networks, which are  multilayer feed-forward neural networks. In the output layer, a linear activation function is utilized, while for the hidden layer neurons here we employ the hyperbolic tangent function. For each atom in the system, the individual values of the ACSF vectors are computed and processed by the atomic neural network corresponding to that element. The resulting atomic energy contributions are then summed to yield the system's potential energy. Forces can be computed as analytic derivatives.

The weight parameters of all atomic neural networks are optimized simultaneously using an iterative gradient-based optimization. In this process, an adaptive, global, extended Kalman filter \cite{kalman1960new,blank1994adaptive} is employed to minimize the errors of the known potential energies and atomic force components for a set of training structures. For more comprehensive details on HDNNPs, including their methodology, training and typical applications, the interested reader is referred to several reviews on this topic~\cite{behler2021four,behler2017first,behler2014representing,behler2015constructing,tokita2023train,omranpour2024perspective}.

\section{Computational Details}\label{sec:computational}
\subsection{Density Functional Theory Calculations}

All electronic structure calculations were performed using the Vienna Ab initio Simulation Package (VASP)~\cite{kresse1996efficient,kresse1996efficiency} within the framework of spin-polarized DFT. The optPBE-vdW functional \cite{perdew1996generalized,klimevs2009chemical,klimevs2011van} was used to treat electronic exchange and correlation effects, and also accounts for van der Waals interactions. Additional on-site Coulomb interactions were included using the DFT+U method to correct for the strongly correlated \(3d\) electrons of cobalt. The effective Hubbard parameter U for Co was set to 2.43~eV, as determined by the method proposed by Dudarev et al.\ \cite{dudarev1998electron}. 

The ionic cores were described by Projector Augmented Wave (PAW) potentials \cite{blochl1994projector}, as derived by Kresse and Joubert \cite{kresse1999ultrasoft,kresse1996efficiency}. The wave functions were expanded in a plane-wave basis set up to a cutoff energy of 500 eV. The Brillouin zone was sampled using a Monkhorst-Pack grid with a \(5 \times 5 \times 5\) k-point mesh. Gaussian smearing with a width of 0.1 eV was applied for partial occupancies, and non-spherical contributions to the PAW spheres were considered. The convergence criterion for electronic self-consistency was set to \(10^{-6}\) eV.

\subsection{Construction of the Reference Data Set}

In order to construct an initial dataset to train a first preliminary set of HDNNPs, \textit{ab initio} molecular dynamics (AIMD) simulations were performed using VASP for a single unit cell. For dealing with the electronic structure within the AIMD simulations, the same settings as described in the previous section were employed, with the only difference being that a \(\Gamma\)-centered \(1 \times 1 \times 1\) k-point grid was used for the Brillouin zone integration to increase the efficiency. The simulations were carried out in the $NVT$ and $NPT$ ensembles, at 300~K and 700~K, with the temperature controlled by a Nosé-Hoover thermostat \cite{P2758}. The simulations were run for approximately 200 ps, using a simulation time step of 2 fs. The first 4000 configurations (8 ps) of the simulation were used to equilibrate the system and discarded. From then on, every 96 steps, configurations were extracted from the AIMD trajectories, and an initial dataset of about 4000 structures was constructed. Then, single-point electronic structure calculations were performed with a dense \(5 \times 5 \times 5\) k-point mesh as mentioned above to compute accurate energies and forces.

Once the initial dataset was prepared and preliminary HDNNPs were constructed, active learning, as implemented by Eckhoff et al.\ \cite{eckhoff2021high,eckhoff2019molecular}, was employed to generate additional structures that exhibit high uncertainty within the ensemble of HDNNPs of similar quality. The HDNNPs were generated using different seeds, which resulted in different train/test splits. Then, electronic structure calculations were performed for these new structures, which were subsequently added to the initial dataset to refine the HDNNP training. Multiple cycles of such active learning were performed until the variance of all trial structures remained close to the RMSEs\cite{tokita2023train}.

\subsection{Construction of the High-Dimensional Neural Network Potential} 

For the construction of the HDNNPs, the RuNNer code~\cite{behler2015constructing,behler2017first} (version from August 16, 2023) was employed. The atomic neural networks of both elements feature three hidden layers with 25, 20, and 15 neurons, respectively. The input neurons correspond to the element-specific ACSFs, and each network has one output node that provides the atomic energy. The ACSF cutoff radius was set to \(R_c = 12 a_0\) (6.35~\AA{}), which is sufficient to provide an accurate description of the atomic interactions in Co$_3$O$_4$. A complete list of the employed atom-centered symmetry functions and their parameters is available in the supporting information.

The HDNNPs were trained on both the DFT energies and the atomic force components of the reference structures. The reference dataset was randomly divided into a training set, comprising approximately 90\% of the structures used for adjusting the weight parameters, and a testing set, consisting of the remaining 10\%, used to evaluate the HDNNP's performance on unknown structures. The Kalman filter parameters were set to \(\lambda = 0.98000\) and \(\nu = 0.99870\). The dataset spans an energy range of 0.2 eV~atom\(^{-1}\) with respect to the optimized Co$_3$O$_4$ crystal structure. Atoms with forces up to 6 eV~\AA{}\(^{-1}\) were included in the training.

\subsection{Molecular Dynamics Simulations}

Molecular dynamics simulations were performed using the Large-scale Atomic/Molecular Massively Parallel Simulator (LAMMPS) (version from 2nd August 2023) \cite{plimpton1995fast}, including the n2p2 library for HDNNPs (version 2.2.0 from 23rd May 2022)\cite{singraber2019parallel}. MD simulations for active learning \cite{eckhoff2021high,eckhoff2019molecular} were run in the $NVT$ and $NVE$ ensembles at various temperatures from 10~K up to 700~K with a time step of \(\Delta t = 1 \, \text{fs}\). The velocity Verlet algorithm was used as the integrator \cite{swope1982computer}. The Nosé–Hoover\cite{nose1984molecular,hoover1985canonical,P2758} thermostat and barostat were used. All $NPT$ simulations were performed at \(p = 1\) bar. The thermodynamic data and trajectories were stored at every step for post-processing analysis.

\subsection{Phonon Calculations}

For the phonon calculations, a Co$_3$O$_4$ unit cell consisting of 56 atoms was initially relaxed using VASP (with DFT) and LAMMPS (with HDNNP) separately. Then, the force constants for the two relaxed structures were derived using the phonopy code (version 2.20.0)\cite{phonopy-phono3py-JPCM,phonopy-phono3py-JPSJ}, based on DFT calculations in VASP and HDNNP calculations via the n2p2 library in LAMMPS. The phonon properties, including phonon band structures and phonon density of states, were then calculated and compared for DFT and HDNNP.

\section{Results and Discussion}\label{sec:results}

\subsection{Density Functional Theory Calculations} \label{subsec:lattParam-DFT}

Since MLPs inherit the quality of the potential energy surface from the underlying electronic structure calculations, validating the DFT setup, and in particular, the employed exchange-correlation functional as the main approximation, is of critical importance. The electronic structure of Co$_3$O$_4$ is difficult to describe, particularly for generalized gradient approximation (GGA) functionals, while more accurate hybrid functionals are computationally much more demanding and thus unfeasible if extended datasets are to be constructed.

\begin{table*}[ht]
\centering
\caption{Lattice constant $a$, band gap, magnetic moment $M$ of Co\textsuperscript{2+}, and the Co--O bond distances (Co\textsuperscript{2+}--O\textsuperscript{2--} and Co\textsuperscript{3+}--O\textsuperscript{2--}) for bulk Co$_3$O$_4$ spinel, with and without Hubbard U correction, using the optPBE-vdW functional. Experimental values and data from other GGA functionals are included for reference. The optPBE-vdW+U (U = 2.43 eV) results, used as the basis for HDNNP construction in this work, are highlighted along with the experimental values.}

\renewcommand{\arraystretch}{1.2} 
\begin{tabular}{|l|>{\centering\arraybackslash}p{1.5cm}|>{\centering\arraybackslash}p{1.5cm}|>{\centering\arraybackslash}p{1.5cm}|>{\centering\arraybackslash}p{1.5cm}|>{\centering\arraybackslash}p{1.5cm}|>{\centering\arraybackslash}p{1.5cm}|}
\hline
 & \small $a$ (\AA{}) & \small band gap (eV) & \small $M$\textsubscript{Co\textsuperscript{2+}} $(\mu_B)$ & \small $d$ (\AA{}) Co\textsuperscript{2+}--O\textsuperscript{2--} & \small $d$ (\AA{}) Co\textsuperscript{3+}--O\textsuperscript{2--} \\
 &  &  &  &  &  \\
\hline
optPBE-vdW & 8.150 & 0.42 & 2.34 & 1.943 & 1.939 \\
\hline
\textbf{optPBE-vdW+U (U = 2.43 eV)} & \textbf{8.156} & \textbf{1.61} & \textbf{2.57} & \textbf{1.955} & \textbf{1.936} \\
\hline
optPBE-vdW+U (U = 3 eV) & 8.157 & 1.82 & 2.62 & 1.957 & 1.935 \\
\hline
PBE\cite{chen2011electronic} & 8.19 & 0.30 & 2.64 & 1.95 & 1.93 \\
\hline
PBE+U\cite{chen2011electronic} & 8.27 & 1.96 & 2.84 & 1.99 & 1.95 \\
\hline
RPBE+U (U = 2.8 eV)\cite{peng2021influence} &  8.23  & 1.57  & 2.64  & 1.98 & 1.95 \\
\hline
PW91+U (U = 3.5 eV)\cite{zasada2015cobalt} & 8.15 & 1.72 & 2.69 & 1.96 & 1.93 \\
\hline
\textbf{Experiment}  & \textbf{8.08\cite{zasada2015cobalt,liu1990high}} & \textbf{1.60\cite{shinde2006supercapacitive}} & \textbf{3.26\cite{roth1964magnetic}} & \textbf{1.94\cite{liu1990high}} & \textbf{1.92\cite{liu1990high}} \\
\hline
\end{tabular}
\label{tab:table1}
\end{table*}

The cubic lattice constant, band gap, magnetic moment of Co\textsuperscript{2+}, the distance between the tetrahedrally coordinated Co ion and the O ion (Co\textsuperscript{2+}--O\textsuperscript{2--}), and the distance between the octahedrally coordinated Co ion and the O ion (Co\textsuperscript{3+}--O\textsuperscript{2--}) for bulk Co$_3$O$_4$ spinel, with and without the Hubbard U correction, are provided for different U values using the optPBE-vdW functional in Table \ref{tab:table1} (the first three rows). Moreover, experimental values and previously published DFT data for several other GGA functionals are included for comparison. Two rows are highlighted, the optPBE-vdW+U (U = 2.43 eV), since it is used as the basis of HDNNP construction in this work, and the experimental values, as the point of reference. We find that the plain optPBE-vdW functional provides a slight overestimation of the lattice constant by about 0.07~\AA{} (0.85~\%), similar to other listed DFT functionals, which sometimes yield even marginally larger values. The marginal improvement of the optPBE-vdW functional with respect to the other GGA functionals might be attributed to the inclusion of the dispersion interactions in the optPBE-vdW functional. However, as expected, all plain GGA functionals strongly underestimate the band gap of Co$_3$O$_4$ and the magnetic moment of Co\textsuperscript{2+}. It is well-known that a Hubbard U correction can be employed to improve the band gap. In this study, a U value of 2.43 eV is calculated by fitting to the experimental band gap of 1.6 eV, with a minor effect (0.006~\AA{}) on the lattice parameter overestimation. As shown in Table \ref{tab:table1}, a Hubbard U value of 3 eV slightly (0.22 eV) overestimates the band gap.

The distance between the tetrahedrally coordinated Co ion and the O ion (Co\textsuperscript{2+}--O\textsuperscript{2--}) and the distance between the octahedrally coordinated Co ion and the O ion (Co\textsuperscript{3+}--O\textsuperscript{2--}) are both overestimated by all GGA functionals, in line with the observation for the lattice constant. However, the fact that the Co\textsuperscript{3+}--O\textsuperscript{2--} distance is $\sim$0.02~\AA{} shorter than the Co\textsuperscript{2+}--O\textsuperscript{2--} distance (due to the higher charge) is consistent between the experimental observation and the optPBE-vdW+U.

Since the reported experimental values for the lattice constant and Co--O distances are extracted at room temperature, and the above DFT calculations correspond to 0~K, the overestimation by GGA is slightly worse than what is shown in the table if thermal effect are included. Apart from the approximate nature of GGA itself, especially for a transition metal oxide such as Co$_3$O$_4$, some other explanation might also be provided: one may argue that the \textit{inclusion/exclusion} of vacancies in the \textit{theoretical/experimental} samples might play a role. In fact, the unit cell studied in this work (and in all other theoretical works mentioned in Table \ref{tab:table1}) represents the \textit{ideal} bulk Co$_3$O$_4$ structure without vacancies. The effect of vacancy concentration on the lattice parameter has been addressed by Koutn{\'a} et al.\cite{koutna2016point}, who studied lattice parameter dependence on the concentration of vacancies in c-MoN and c-TaN supercells based on DFT calculations. Their calculations showed a strong decrease in lattice parameters as vacancy concentration increased, with a reduction of up to 0.2~\AA{} as vacancy concentration increased from 0\% to 30\%! This observation appears to be general, as it has also been reported by Stampfl and Freeman \cite{stampfl2003metallic} and Grumski et al.\cite{grumski2013ab} for c-TaN, and by Lowther \cite{lowther2004lattice} for c-MoN.

In summary, the inclusion of vacancies might be essential for better predicting experimental observables. However, this presents two main challenges: one related to DFT and the other to the structure of Co$_3$O$_4$. The first challenge involves the system size, which must be small enough to be manageable by DFT in terms of computational cost but large enough to represent realistic vacancy concentrations and distributions within the bulk system. The second challenge pertains to the magnetic moment of Co\textsuperscript{2+} ions and possibly also interconversion between Co\textsuperscript{2+} and Co\textsuperscript{3+} for charge neutrality. Removing any of the three constituent ions of Co$_3$O$_4$ (i.e., Co\textsuperscript{2+}, Co\textsuperscript{3+}, and O\textsuperscript{2-}) and understanding its effect on the magnetic properties of Co$_3$O$_4$ is a highly complex task beyond the scope of this work. However, HDNNPs could be a promising approach to address the first issue, enabling simulations on a larger scale and thus including more vacancies in a larger system. Additionally, HDNNPs could be useful for studying the thermal effects on vacancy migration and its related consequences. The complexity and errors of the electronic structure calculations and their implications for constructing MLPs/HDNNPs will be addressed at the end of the discussion section (see Section \ref{subsubsec:time-length}).

We note that there is also some uncertainty in the experimental data in Table~\ref{tab:table1}, and other values have also been reported for both the band gap and lattice parameter of Co$_3$O$_4$. For example, in a study by Flechas et al.\cite{cardenas2021effect}, it was shown how the calcination temperature during sample preparation affects the behavior of Co$_3$O$_4$ nanoparticles, including the lattice parameter. However, since the experimental values mentioned in Table \ref{tab:table1} are the ones most frequently referred to in the literature, we have chosen them as our reference here. Overall, as the optPBE-vdW+U functional with U = 2.43~eV provides a reasonable description of both the geometric and electronic structure of Co$_3$O$_4$, we will use this setup as the reference method to compute the dataset for the HDNNP construction.

\subsection{High-Dimensional Neural Network Potential}

After completing the iterative generation of the reference dataset by active learning, a total of 10,766 bulk Co$_3$O$_4$ structures (each containing 56 atoms) were available, providing 10,766 potential energies and 602,896 force components (100\% of all energies and 10\% of all forces were used). From this dataset, 9,692 structures were used for training the HDNNP, and 1,074 structures were used as an independent test set. The root mean squared error (RMSE) achieved for the energy is 0.409 meV/atom for the training set and 0.434 meV/atom for the test set. The RMSE for the force components is 0.055 eV/\AA{} for the training set and 0.056 eV/\AA{} for the test set. Both the RMSEs for the energy and forces are in the typical order of magnitude of state-of-the-art MLPs.

\begin{figure}[H]
\centering
\begin{subfigure}[b]{0.49\textwidth}
    \centering
    \includegraphics[width=\textwidth, trim= 0 0 0 0, clip=true]{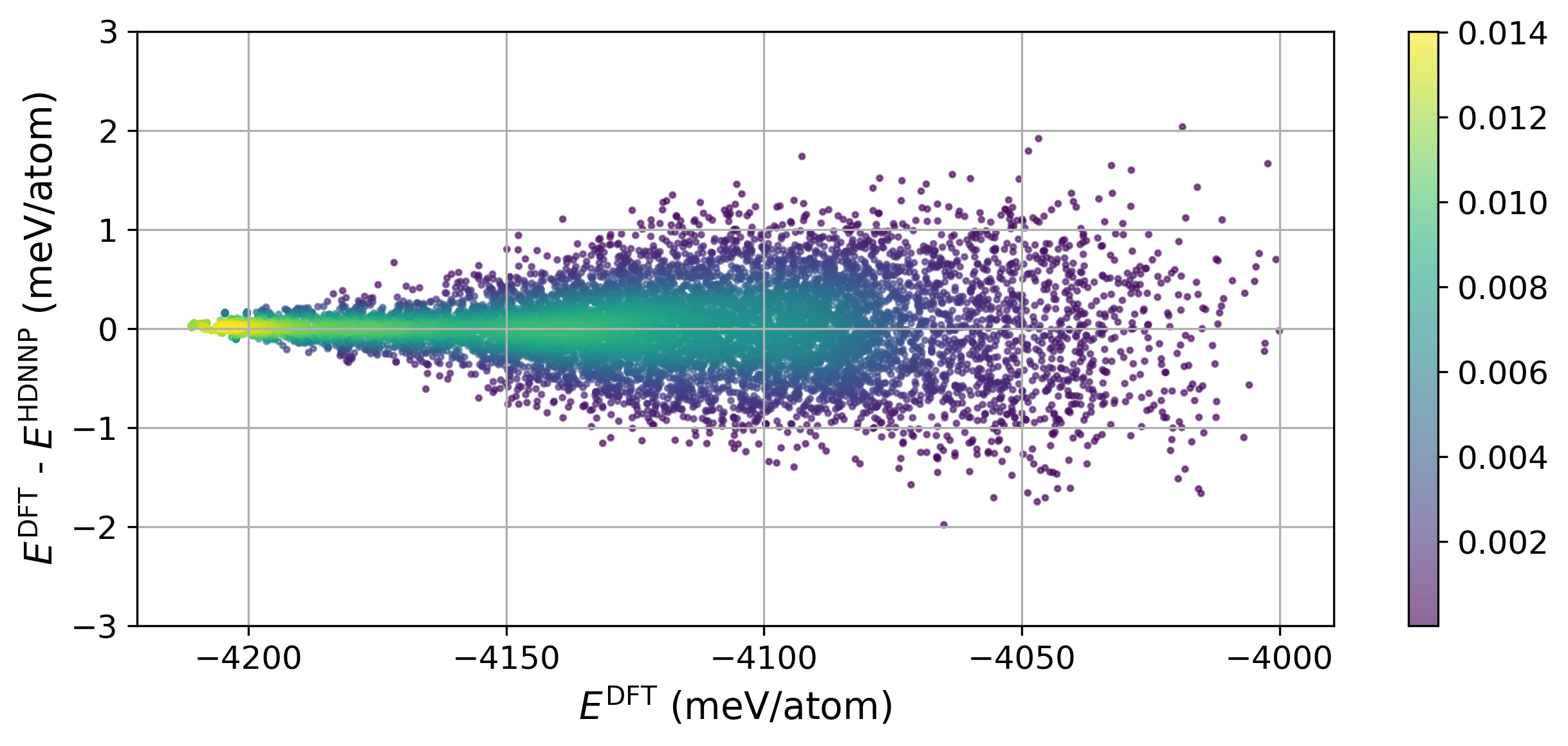}
    \caption{}
    \label{fig:dE_train}
\end{subfigure}
\hfill
\begin{subfigure}[b]{0.49\textwidth}
    \centering
    \includegraphics[width=\textwidth, trim= 0 0 0 0, clip=true]{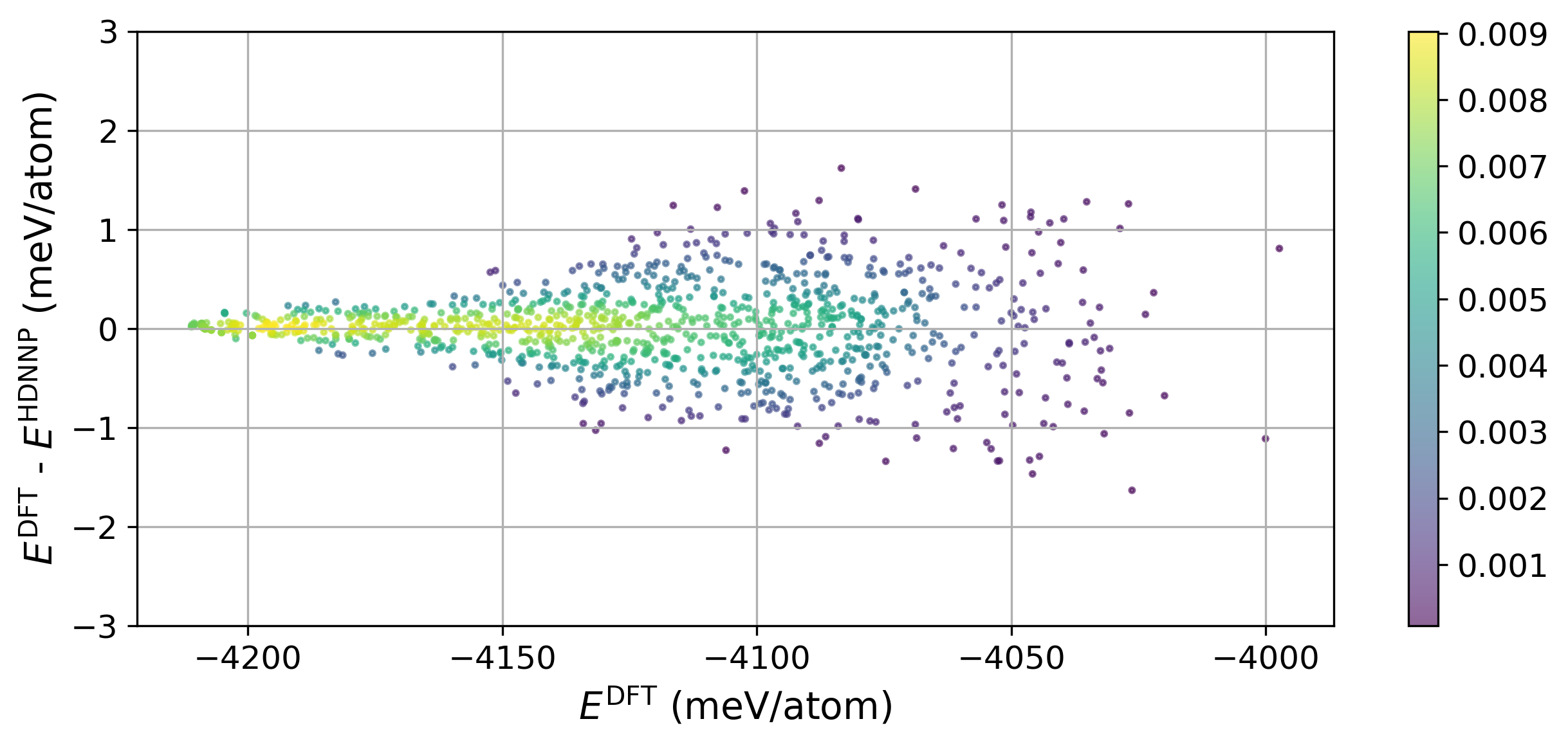}
    \caption{}
    \label{fig:dE_test}
\end{subfigure}
\caption{Energy errors between the HDNNP predictions and DFT results for (a) the training dataset and (b) the testing dataset. Even for outliers, the errors are consistently less than 2 meV/atom across both datasets. The data points are color-coded based on their relative density, highlighting regions of higher data population.}
\label{fig:dE}
\end{figure}

\begin{figure}[h]
\centering
\begin{subfigure}[b]{0.49\textwidth}
    \centering
    \includegraphics[width=\textwidth, trim= 0 0 0 0, clip=true]{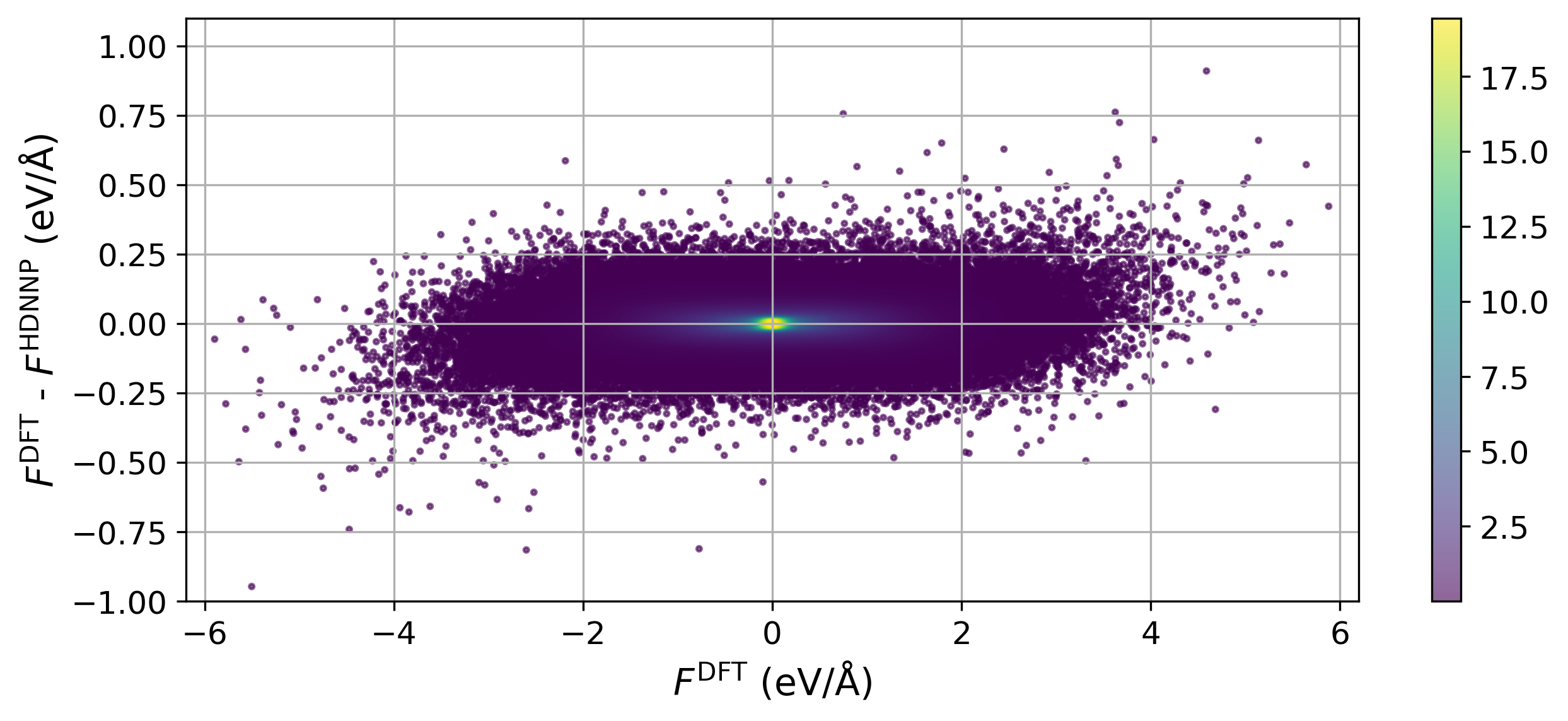}
    \caption{}
    \label{fig:dF_train}
\end{subfigure}
\hfill
\begin{subfigure}[b]{0.49\textwidth}
    \centering
    \includegraphics[width=\textwidth, trim= 0 0 0 0, clip=true]{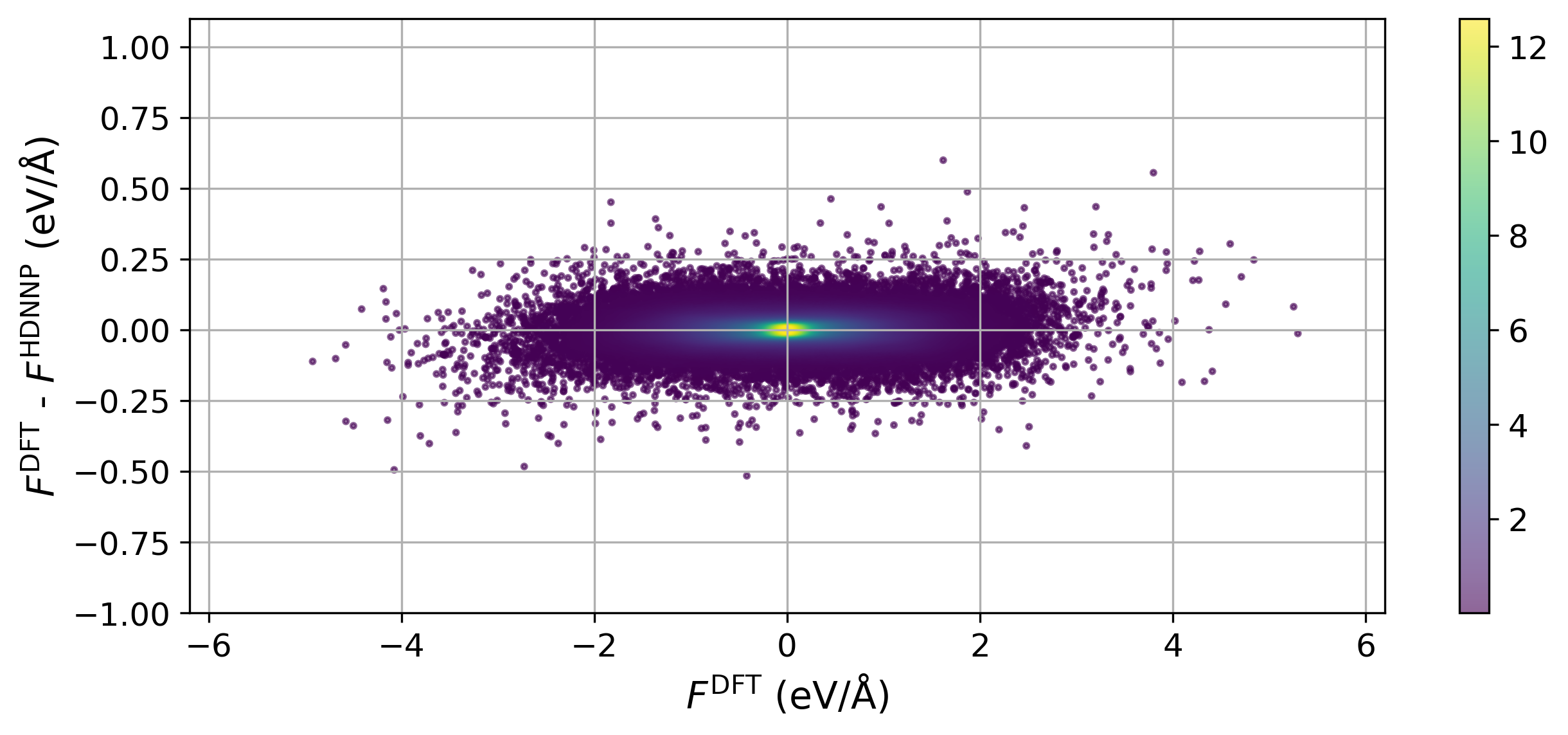}
    \caption{}
    \label{fig:dF_test}
\end{subfigure}
\caption{Force component errors between the HDNNP predictions and DFT results for (a) the training dataset and (b) the testing dataset. The data points are color-coded based on their relative density, with brighter regions indicating areas of higher data population.}
\label{fig:dF}
\end{figure}

As can be seen in Figure \ref{fig:dE}, the energy deviations between the HDNNP predictions and the DFT reference values are generally less than 2 meV/atom and for most points substantially smaller, and the distribution of the test data is very similar to the training data. The lower energy structures exhibit the highest accuracy. A similar observation can be made for the force component errors (Figure \ref{fig:dF}). Here, even the largest force errors are always less than 1 eV/\AA{}. The center of the data, where the DFT data density is highest also corresponds to the region with the least HDNNP force prediction error. Overall, Figures \ref{fig:dE} and \ref{fig:dF} show that the low energy and force errors of the reference data satisfy the necessary, although not sufficient, conditions for a high-quality HDNNP. Further validation requires the computation of static and dynamic properties employing the potential.

Like all MLPs, HDNNPs have a limited transferability beyond the dataset that has been used in its construction. The range of validity of the current potential for bulk Co$_3$O$_4$ spinel spans temperatures from 0~K to 700~K, which is sufficient for applications in catalysis where experiments are rarely conducted above 450~K. The reason for this is that beyond this temperature, most adsorbed species would desorb from the surface~\cite{omranpoor20232}. 

\subsection{Static Properties}

\begin{figure}[H]
\centering
\includegraphics[width=0.48\textwidth, trim= 0 0 0 0, clip=true]{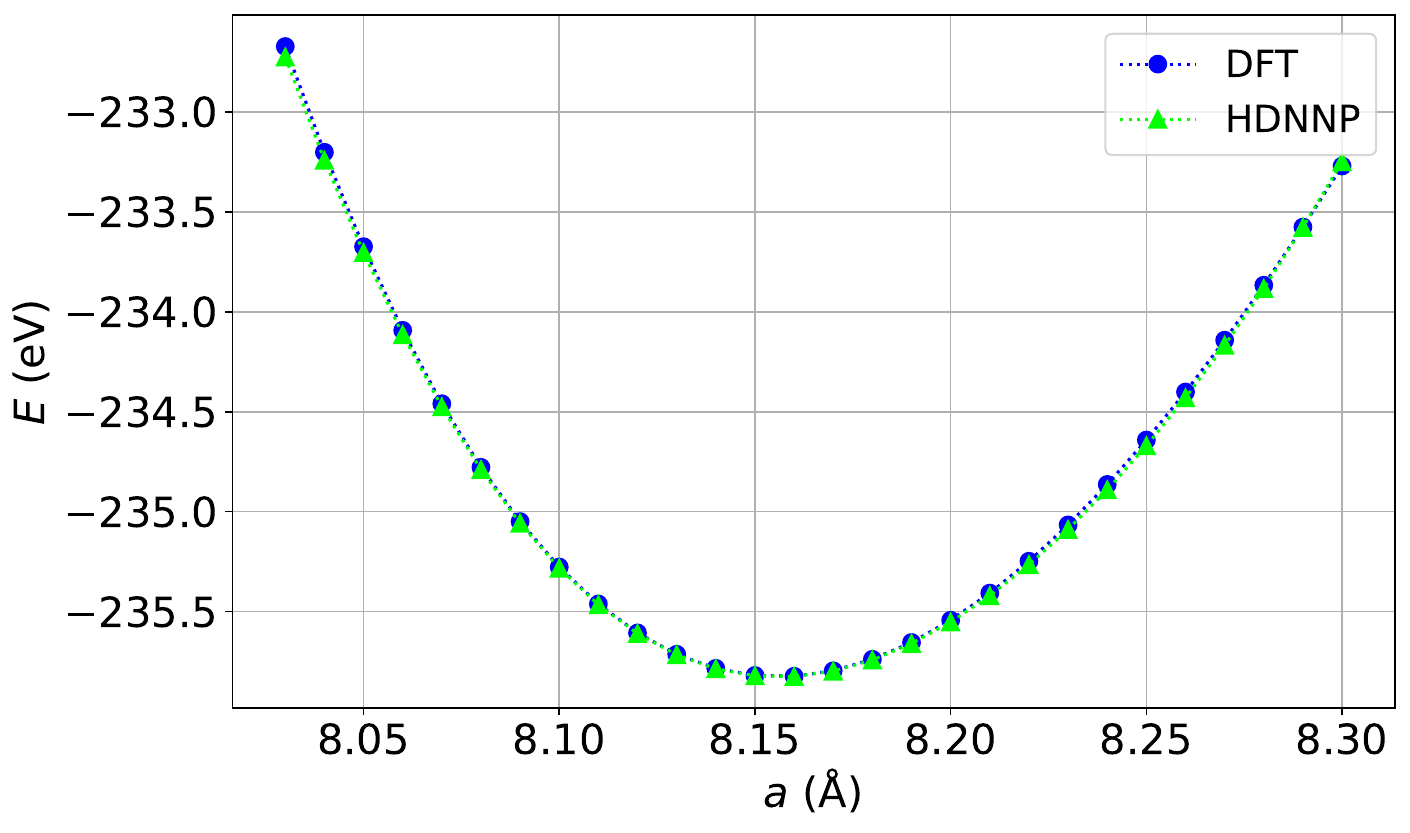}
\caption{Total energy (\textit{E}) versus lattice parameter (\textit{a}) of bulk Co$_3$O$_4$ spinel for the HDNNP and DFT. The predicted lattice parameters using the Birch-Murnaghan equation of state are 8.1561~\AA{} for DFT and 8.1569~\AA{} for the HDNNP.}
\label{fig:E-V}
\end{figure}

Figure \ref{fig:E-V} presents a plot of the total energy \textit{E} versus the lattice parameter \textit{a} value for two datasets obtained with the HDNNP and the DFT calculations. As shown in the figure, there is excellent agreement between the two, particularly near the energy minimum. By employing the Birch-Murnaghan equation of state\cite{birch1947finite,murnaghan1944compressibility,poirier2000introduction} to the data, the DFT and HDNNP lattice parameters have been determined as 8.1561~\AA{} and 8.1569~\AA{}, respectively. Thus, the HDNNP prediction error for the lattice constant is less than 0.01\% compared to the DFT method on which it was trained, which is much smaller than the deviation between DFT and experiment such that the HDNNP does not introduce a significant additional error to the DFT uncertainty.

Table \ref{tab:elastic} provides all the elastic constants, including the components of the elastic constant tensor $C_{ij}$, the bulk modulus $K$, shear modulus $G$, and Poisson's ratio $\nu$, calculated using the method suggested by Clavier and Thompson \cite{clavier2023computation}. Unfortunately, the only available component in the literature for comparison is the bulk modulus, both experimentally and theoretically. The experimental estimate of the bulk modulus is about $190 \pm 5$ GPa \cite{bai2012charge}, and other DFT studies calculated it to be 192 GPa \cite{chen2011electronic} using PBE+U. Both values agree very well with the HDNNP prediction of 190.3 GPa ($K$ in Table \ref{tab:elastic}). Due to the computational cost of performing the aforementioned (Clavier and Thompson) method with DFT, a cheaper approximation of calculating the bulk modulus from the EOS, as implemented in the Atomic Simulation Environment (ASE) \cite{larsen2017atomic}, is also provided, estimating the value to be 188 GPa for the optPBE-vdW+U.

\begin{table}[H]
    \caption{Bulk properties of Co$_3$O$_4$ spinel, including the components of the elastic constant tensor $C_{ij}$, bulk modulus $K$, shear modulus $G$, and Poisson's ratio $\nu$. All units are in GPa except for the unitless Poisson's ratio.}
    \label{tab:elastic}
    \centering
    \renewcommand{\arraystretch}{1.3} 
    \begin{tabular}{|>{\centering\arraybackslash}m{3.6cm}|>{\centering\arraybackslash}m{1.7cm}|>{\centering\arraybackslash}m{2.2cm}|}
        \hline
        \small{Property} & \small{Symbol} & \small{HDNNP} \\
        \hline
        \multirow{21}{*}{Elastic Constants (GPa)} & $C_{11}$ & 267.9 \\
        & $C_{22}$ & 267.9 \\
        & $C_{33}$ & 267.9 \\
        & $C_{12}$ & 151.6 \\
        & $C_{13}$ & 151.6 \\
        & $C_{23}$ & 151.6 \\
        & $C_{44}$ & 110.6 \\
        & $C_{55}$ & 110.6 \\
        & $C_{66}$ & 110.6 \\
        & $C_{14}$ & $2.7 \times 10^{-7}$ \\
        & $C_{15}$ & $-8.8 \times 10^{-7}$ \\
        & $C_{16}$ & $-6.3 \times 10^{-7}$ \\
        & $C_{24}$ & $9.3 \times 10^{-7}$ \\
        & $C_{25}$ & $-3.4 \times 10^{-7}$ \\
        & $C_{26}$ & $-2.2 \times 10^{-6}$ \\
        & $C_{34}$ & $-7.8 \times 10^{-7}$ \\
        & $C_{35}$ & $-4.6 \times 10^{-7}$ \\
        & $C_{36}$ & $1.3 \times 10^{-6}$ \\
        & $C_{45}$ & $-8.3 \times 10^{-7}$ \\
        & $C_{46}$ & $3.9 \times 10^{-7}$ \\
        & $C_{56}$ & $1.5 \times 10^{-7}$ \\
        \hline
        \textbf{Bulk Modulus (GPa)} & \textbf{$K$} & \textbf{190.3} \\
        Shear Modulus 1 (GPa) & $G_1$ & 110.6 \\
        Shear Modulus 2 (GPa)& $G_2$ & 58.1 \\
        Poisson's Ratio & $\nu$ & 0.4 \\
        \hline
    \end{tabular}
\end{table}

\subsection{Vibrational Properties}

\begin{figure}[H]
\centering
\begin{subfigure}{0.4\textwidth}
    \centering
    \caption{Phonon Band Structures}
    \label{fig:phonon_bands}
    \includegraphics[width=\textwidth, trim= 10 10 60 65, clip=true]{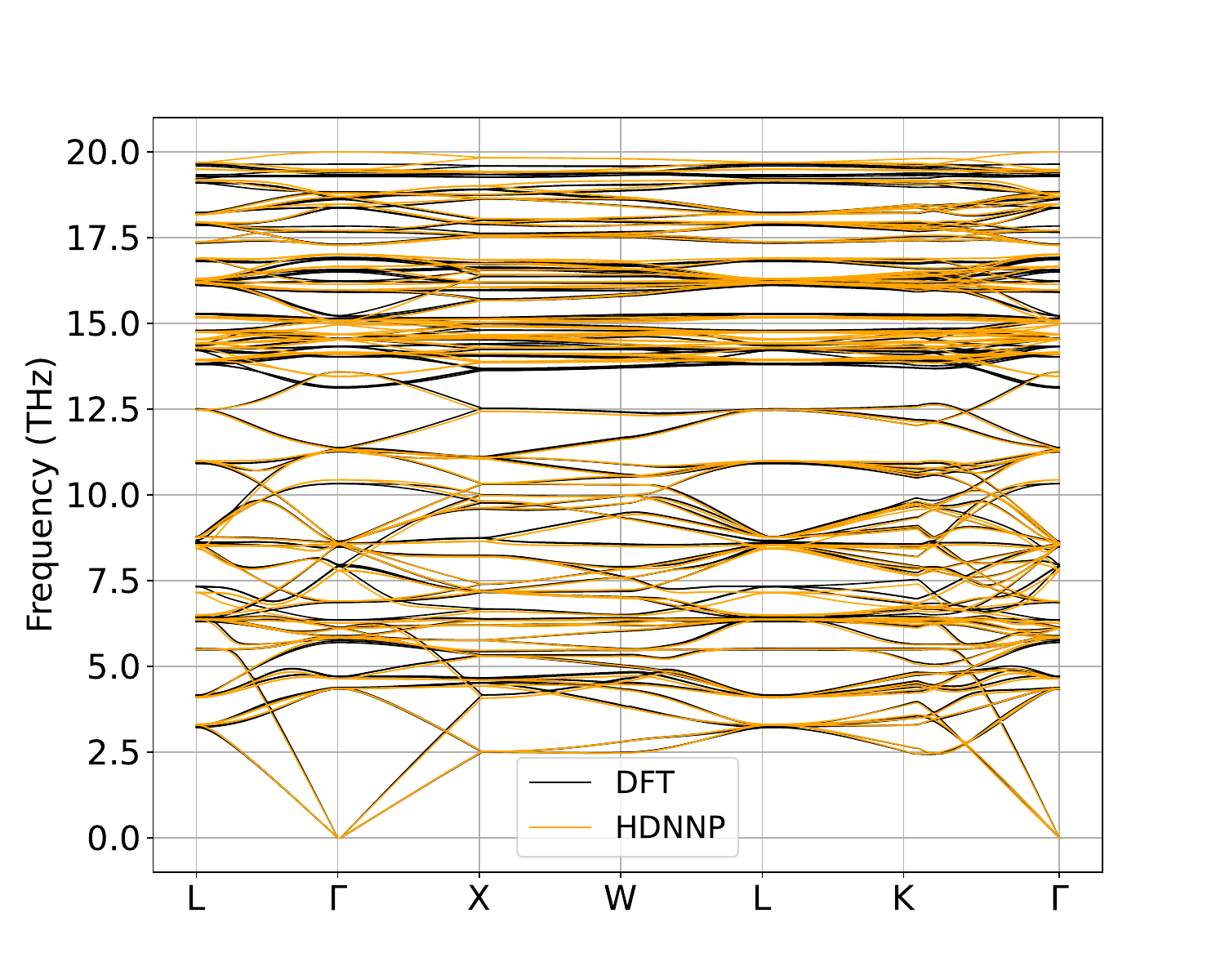}
\end{subfigure}%
\hfill
\begin{subfigure}{0.4\textwidth}
    \centering
    \caption{Density of States (DOS)}
    \label{fig:dos_comparison}
    \includegraphics[width=\textwidth, trim= 10 10 60 65, clip=true]{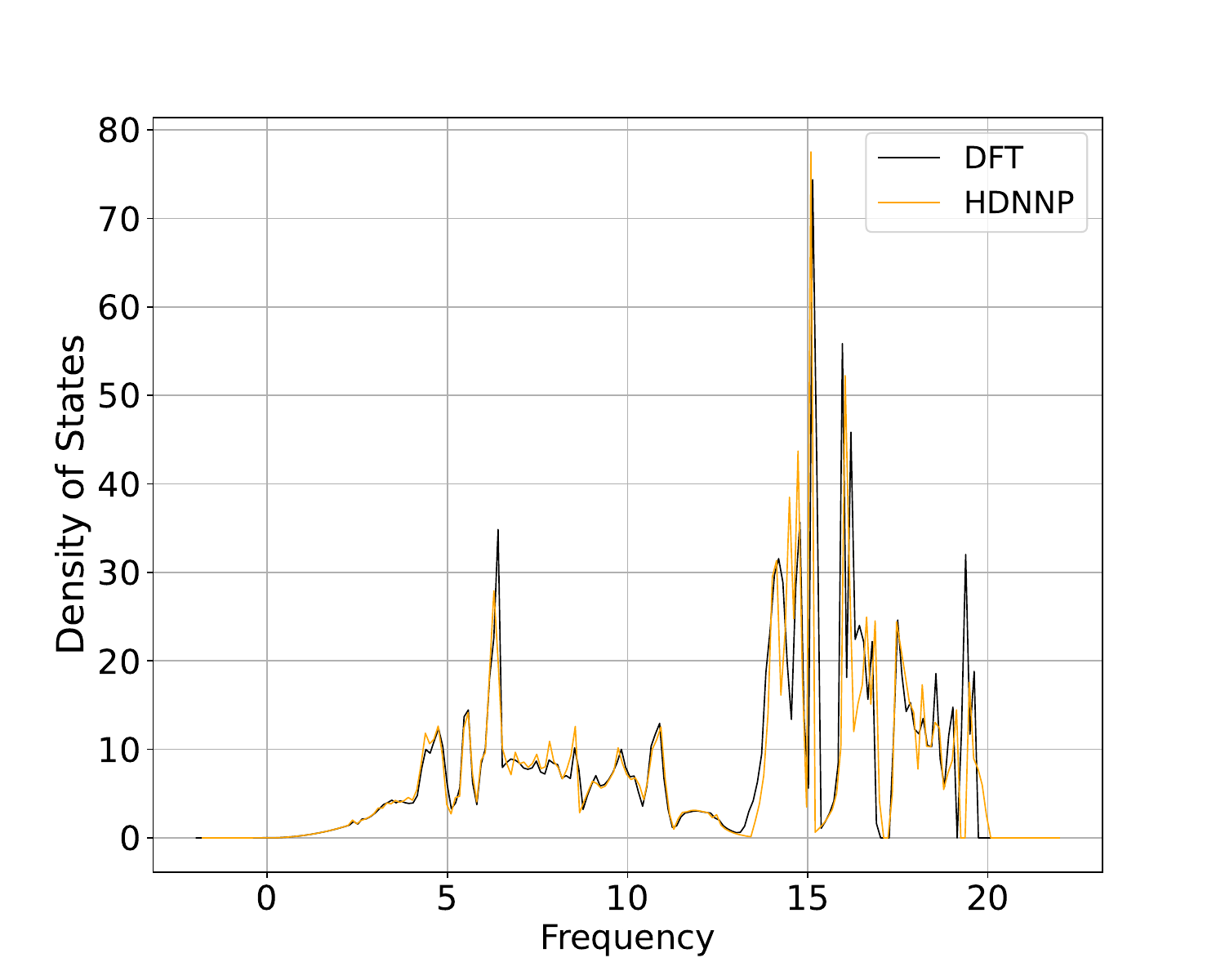}
\end{subfigure}%
\hfill
\begin{subfigure}{0.4\textwidth}
    \centering
    \caption{Thermal Properties}
    \label{fig:thermal_properties}
    \includegraphics[width=\textwidth, trim= 10 10 60 65, clip=true]{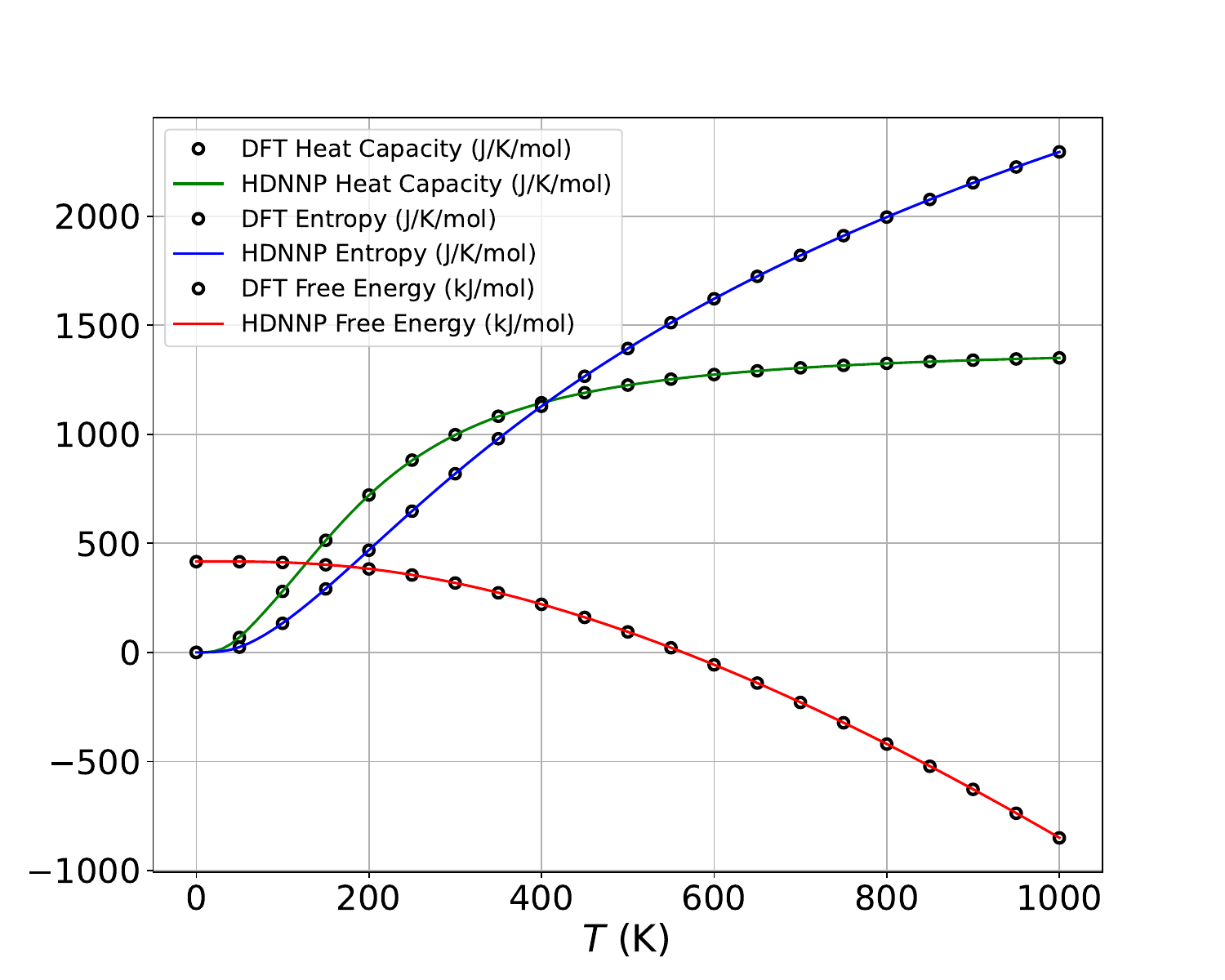}
\end{subfigure}
\caption{Phonon properties of bulk Co$_3$O$_4$ spinel calculated using force constants\cite{togo2015first,togo2010first} from DFT and HDNNP calculations. Panel (a) shows the phonon band structures for DFT (black) and the HDNNP (orange). In (b) the phonon density of states (DOS) obtained from DFT (black) and HDNNP (orange) is plotted. Panel (c) displays the thermal properties derived from phonon calculations using HDNNP, including heat capacity (green), entropy (blue), and free energy (red), compared with DFT (black circles).}
\label{fig:phonons}
\end{figure}

Figure \ref{fig:phonons} shows a comparison of various DFT and HDNNP phonon properties. The phonon band structures in panel (a) show an excellent overall agreement. In the low-frequency regions they match almost perfectly, while exhibiting only small differences in the high-frequency regions, especially in the frequency range between 13 and 15 THz. The overall band structures match closely resulting in very similar phonon excitation properties and thermal energy transfer mechanisms of the system.

Since the acoustic modes begin at zero frequency at the Gamma point, this indicates that there are no forces causing spontaneous vibrations when the atoms are at rest. The phonon band structure also suggests that the bulk Co$_3$O$_4$ spinel is dynamically stable due to the absence of negative (or imaginary) frequencies throughout the phonon band structure, which would indicate instabilities leading to structural changes or phase transitions.

Recovering the phonon density of states (DOS) using the HDNNP is crucial, as the DOS is highly sensitive to small changes in the potential energy surface, which can significantly alter energy levels or phonon modes. Figure \ref{fig:phonons} (b) shows the phonon density of states obtained from DFT and HDNNP calculations. Overall, the distribution of states across the frequency spectrum is very similar, particularly in the low-frequency range, and both methods predict the same key features in the DOS, such as the position and intensity of major peaks. This indicates that the HDNNP is able to capture the main vibrational characteristics of the material very accurately when compared to DFT calculations. However, the only major discrepancies between DFT and HDNNP are observed at peaks around the 15 THz frequency range, in line with our observations of the phonon band structure.

Figure \ref{fig:phonons} (c) illustrates the thermal properties from the phonon calculations, including heat capacity, entropy, and free energy. In general, bulk Co$_3$O$_4$ spinel appears to be thermodynamically stable across the temperature range, as indicated by the continuous decrease in free energy with increasing temperature. The Co$_3$O$_4$ entropy increases with temperature, which is typical and suggests no unusual phase transitions within the temperature range analyzed. Thus, HDNNP and DFT thermal properties also match perfectly. It should be noted that Figure \ref{fig:phonons} (c) is not directly extracted from MD simulation at finite temperature, but rather from the phonon properties. For more information on how such properties are calculated using the force constants, see  Refs.~\citenum{togo2015first,togo2010first}. 

In summary, it can be concluded that the HDNNP is able to reproduce phonon properties of the bulk Co$_3$O$_4$ spinel in excellent agreement with the DFT phonon calculations.

\subsection{Molecular Dynamics Simulations} 
\subsubsection{The Impact of Temperature} \label{subsubsec:Finite-T}

\begin{figure}[H]
\centering
\includegraphics[width=0.48\textwidth, trim= 10 10 10 10, clip=true]{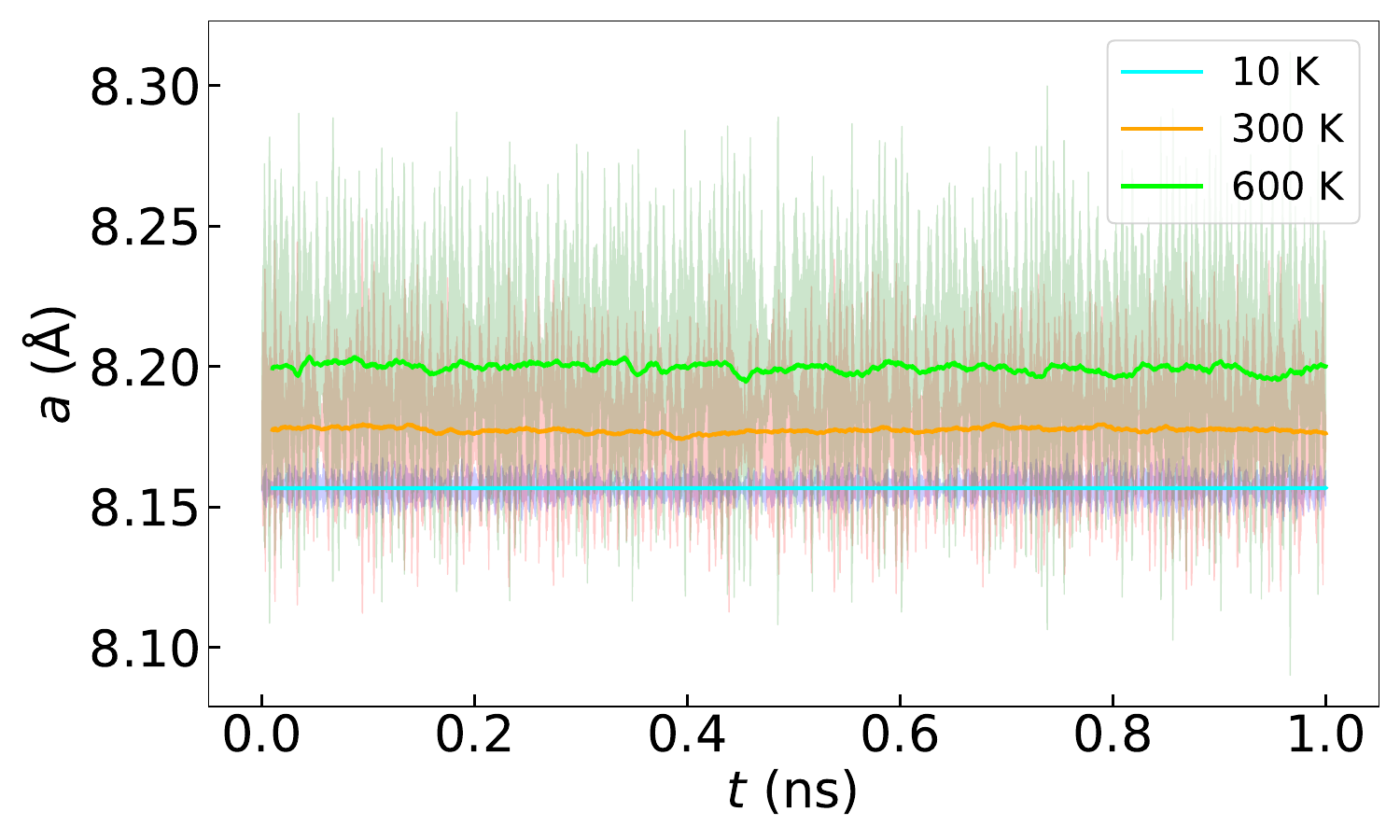}
\caption{Evolution of the lattice parameter (\textit{a}) during 1~ns MD simulations in the $NPT$ ensemble at 10~K, 300~K, and 600~K for a single unit cell. The moving averages (over 10,000-step windows) are depicted in blue, orange, and green for 10~K, 300~K, and 600~K, respectively. The results demonstrate the structural stability of the HDNNP-driven simulations over the entire 1~ns runtime, with the lattice parameters showing a slight increase due to thermal expansion. Specifically, the lattice parameter increases from 8.157 \AA\ at 10~K to 8.177 \AA\ at 300~K, and to 8.200 \AA\ at 600~K. This thermal expansion is consistent with experimental data reported by Liu et al.\cite{liu1990high}, further validating the accuracy of the HDNNP simulations.}

\label{fig:a-t}
\end{figure}

Figure \ref{fig:a-t} shows the evolution of the lattice parameter \textit{a} during 1~ns MD simulations in the $NPT$ ensemble at 10~K, 300~K, and 600~K for a single unit cell. The moving averages (over 10,000-step windows) are also provided and shown in blue, orange, and green for the aforementioned temperatures, respectively. Overall, the trajectories provide evidence for the structural stability of the HDNNP-driven simulations over the entire runtime of 1~ns for the system up to the maximum investigated temperature of 600~K. It was mentioned that the DFT lattice parameter predicted by the underlying optPBE+U calculation at 0~K is 8.156 \AA\ (see Section \ref{subsec:lattParam-DFT}). At 10~K, the average value of the lattice parameter is approximately 8.157 \AA, 8.177 \AA\ at 300~K, and 8.200 \AA\ at 600~K.

It can thus be deduced that heating the bulk Co$_3$O$_4$ spinel from 0~K to 300~K results in a 0.014 \AA\ increase, and from 300~K to 600~K results in a 0.023 \AA\ increase in the lattice parameter due to the thermal expansion of the material. For the latter observation, there is experimental data available from Liu et al.\cite{liu1990high} reporting a thermal expansion of 0.026 \AA\ when the temperature increases from 296~K to 773~K, which is in excellent agreement with the HDNNP results, showcasing how the thermal effect can be accurately captured by the HDNNP-driven MD simulations.

\begin{figure}[H]
\centering
\begin{subfigure}{0.46\textwidth}
    \centering
    \caption{Co-Co}
    \label{fig:Co-Co_RDF}
    \includegraphics[width=\textwidth, trim=  0 0 0 0, clip=true]{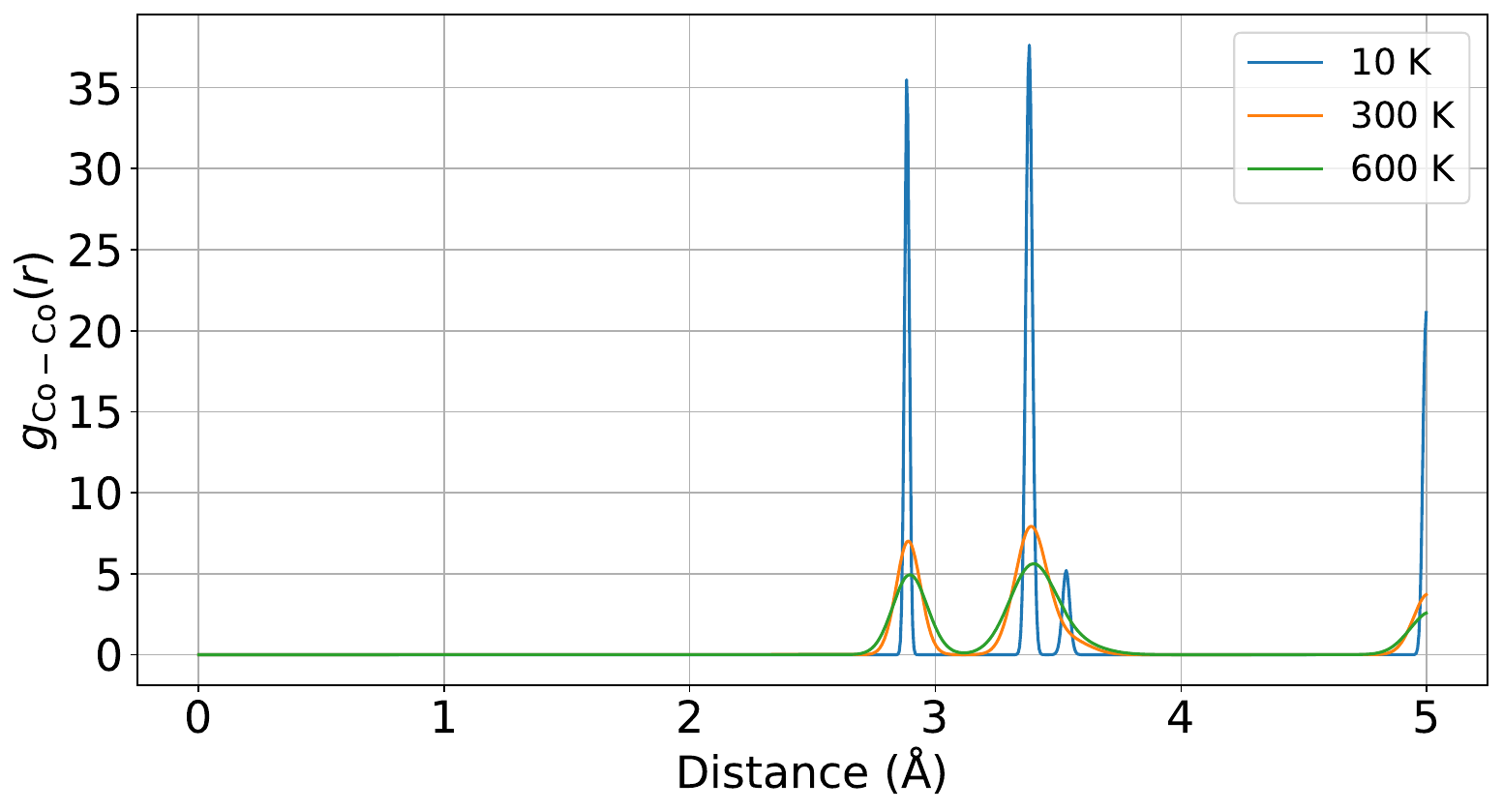}
\end{subfigure}%
\hfill
\begin{subfigure}{0.46\textwidth}
    \centering
    \caption{O-O}
    \label{fig:O-O_RDF}
    \includegraphics[width=\textwidth, trim= 0 0 0 0, clip=true]{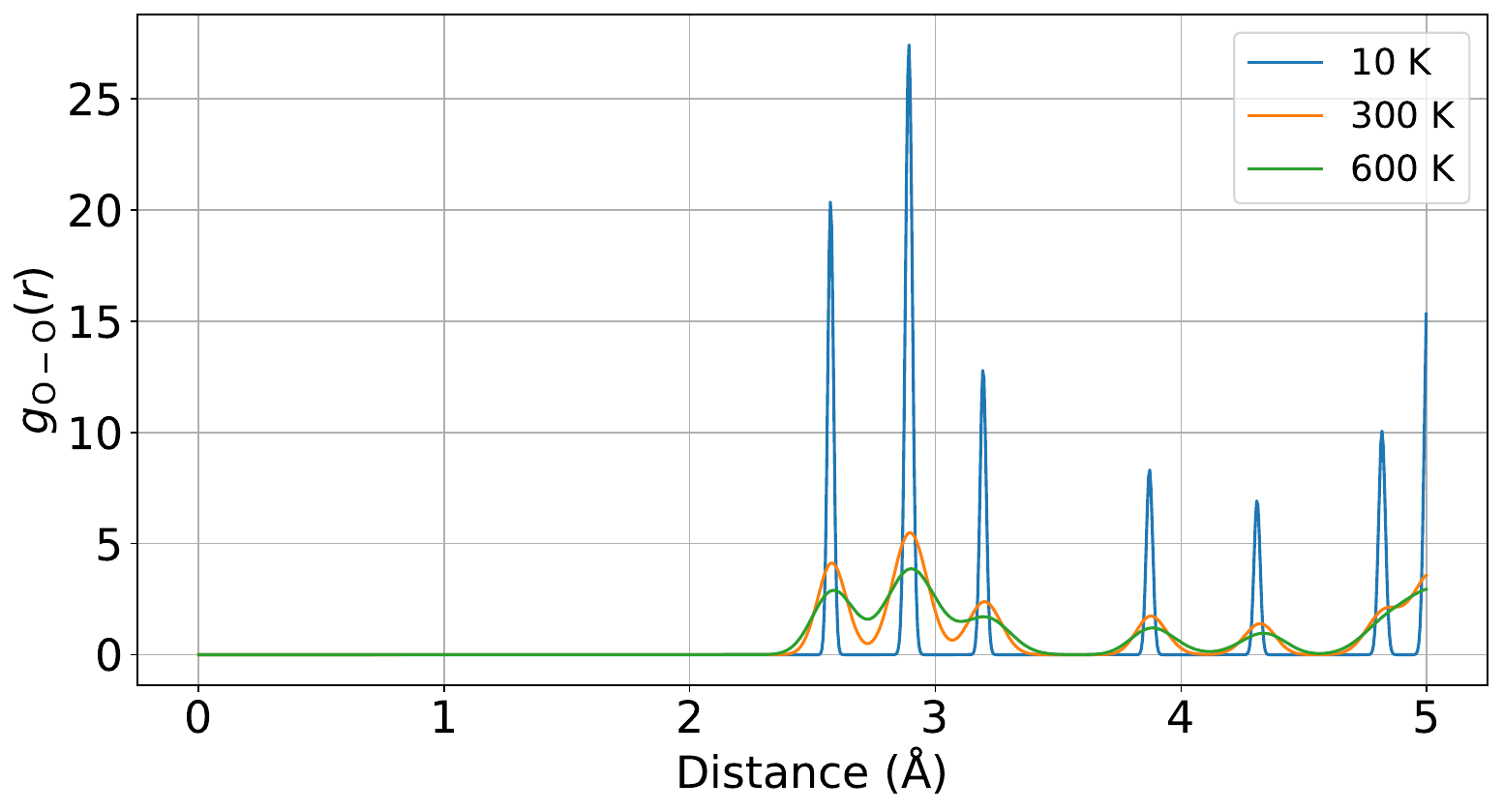}
\end{subfigure}%
\hfill
\begin{subfigure}{0.46\textwidth}
    \centering
    \caption{Co-O}
    \label{fig:Co-O_RDF}
    \includegraphics[width=\textwidth, trim= 0 0 0 0, clip=true]{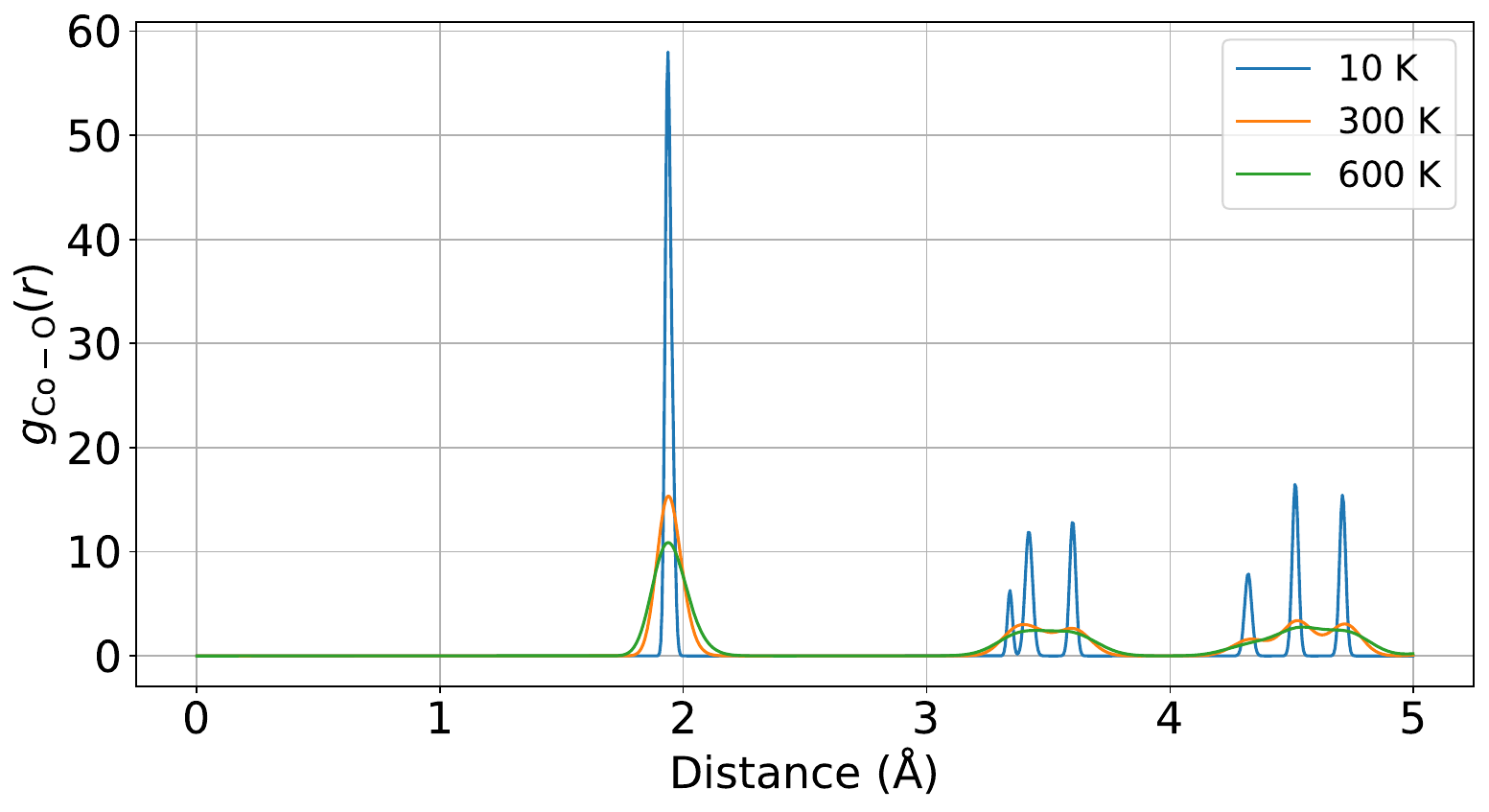}
\end{subfigure}
\caption{Radial distribution functions (RDFs) from the three $NPT$ simulations at 10~K, 300~K, and 600~K. (a) Co--Co interatomic distances: Sharp peaks are observed at 10~K, with broadening at higher temperatures due to increased thermal motion. The first peak at 2.88~\AA{} corresponds to the distance between pairs of Co\textsuperscript{3+}, and the second peak at 3.38~\AA{} represents the distance between pairs of Co\textsuperscript{3+} and Co\textsuperscript{2+} ions. The third peak around 2.50~\AA{}, visible only at 10~K, represents the distance between pairs of Co\textsuperscript{2+} ions. As thermal motion increases at 300~K and 600~K, this peak merges with the second peak. (b) O--O interatomic distances: The first three peaks at 2.57~\AA{}, 2.89~\AA{}, and 3.19~\AA{} correspond to the closest oxygen atoms, with substantial broadening as temperature increases. (c) Co--O interatomic distances: The first peak at 1.96~\AA{} represents the most probable distance to find an O atom around a Co ion. The Co\textsuperscript{3+}--O\textsuperscript{2--} and Co\textsuperscript{2+}--O\textsuperscript{2--} bond lengths are averaged out due to thermal motion.}
\label{fig:RDFs}
\end{figure}

Since we now discuss the interatomic distances of different coordination environments, the naming convention introduced in section \ref{sec:introduction} will be employed here. Figure \ref{fig:RDFs} shows the radial distribution functions (RDFs) of the three $NPT$ simulations. Panel (a) illustrates the Co--Co interatomic distances, where sharp peaks are observed in the 10~K simulation, characteristic of solids at low temperatures. These peaks broaden at 300~K and even more so at 600~K, indicating increased thermal motion with rising temperature. The first peak, representing the distance between the pairs of octahedrally coordinated Co ions (Co\textsuperscript{3+}), occurs around 2.88~\AA{}, while the second peak, representing the distance between pairs consisting of one octahedrally (Co\textsuperscript{3+}) and one tetrahedrally (Co\textsuperscript{2+}) coordinated Co ion, is located at about 3.38~\AA{}. The third peak around 2.50~\AA{}, visible only at 10~K, represents the distance between the pairs of tetrahedrally coordinated Co ions (Co\textsuperscript{2+}). As the thermal motion increases at elevated temperatures, these ions become more mobile, causing the second and third peaks to merge at 300~K and 600~K.

Figure \ref{fig:RDFs}~(b) shows the O--O interatomic distances. The first three peaks correspond to the closest oxygen atoms at 2.57~\AA{}, 2.89~\AA{}, and 3.19~\AA{}, respectively. As the temperature increases, the distributions broaden substantially. Figure \ref{fig:RDFs}~(c) depicts the Co--O interatomic distances. The first peak, which represents the most probable distance to find an O atom around a Co ion, occurs at 1.96~\AA{}. As discussed in Section \ref{subsec:lattParam-DFT}, the Co\textsuperscript{3+}--O\textsuperscript{2--} and Co\textsuperscript{2+}--O\textsuperscript{2--} bond lengths are approximately 1.95~\AA{} and 1.96~\AA{}, respectively. Since these peaks are no longer distinct in this RDF, it can be concluded that they are averaged out due to thermal motion and the close proximity of these two values.

Overall, except for the broadening of the major peaks due to thermal motion, there are no significant shifts in the peak locations. In the experiments by Liu et al.\cite{liu1990high}, a similar trend was observed, with a reported increase of 0.004~\AA{} for the Co\textsuperscript{2+}--O\textsuperscript{2--} distance and an increase of 0.008~\AA{} for the Co\textsuperscript{3+}--O\textsuperscript{2--} distance when the temperature increased from 296~K to 773~K, which is in line with our observation.

One of the most important thermal properties that can be derived from $NPT$ simulations is the linear thermal expansion coefficient \(\alpha\), which quantifies how the lattice parameter changes with temperature. It can be calculated as
\begin{equation}
\alpha = \frac{1}{a_0} \frac{da}{dT} \quad, \label{eq:thermalexpansion}
\end{equation}
where \(a_0\) is the lattice parameter at the reference temperature, and \(\frac{da}{dT}\) is the slope of the lattice parameter versus the temperature curve. 

\begin{figure}[H]
\centering
\includegraphics[width=0.48\textwidth, trim= 0 0 0 0, clip=true]{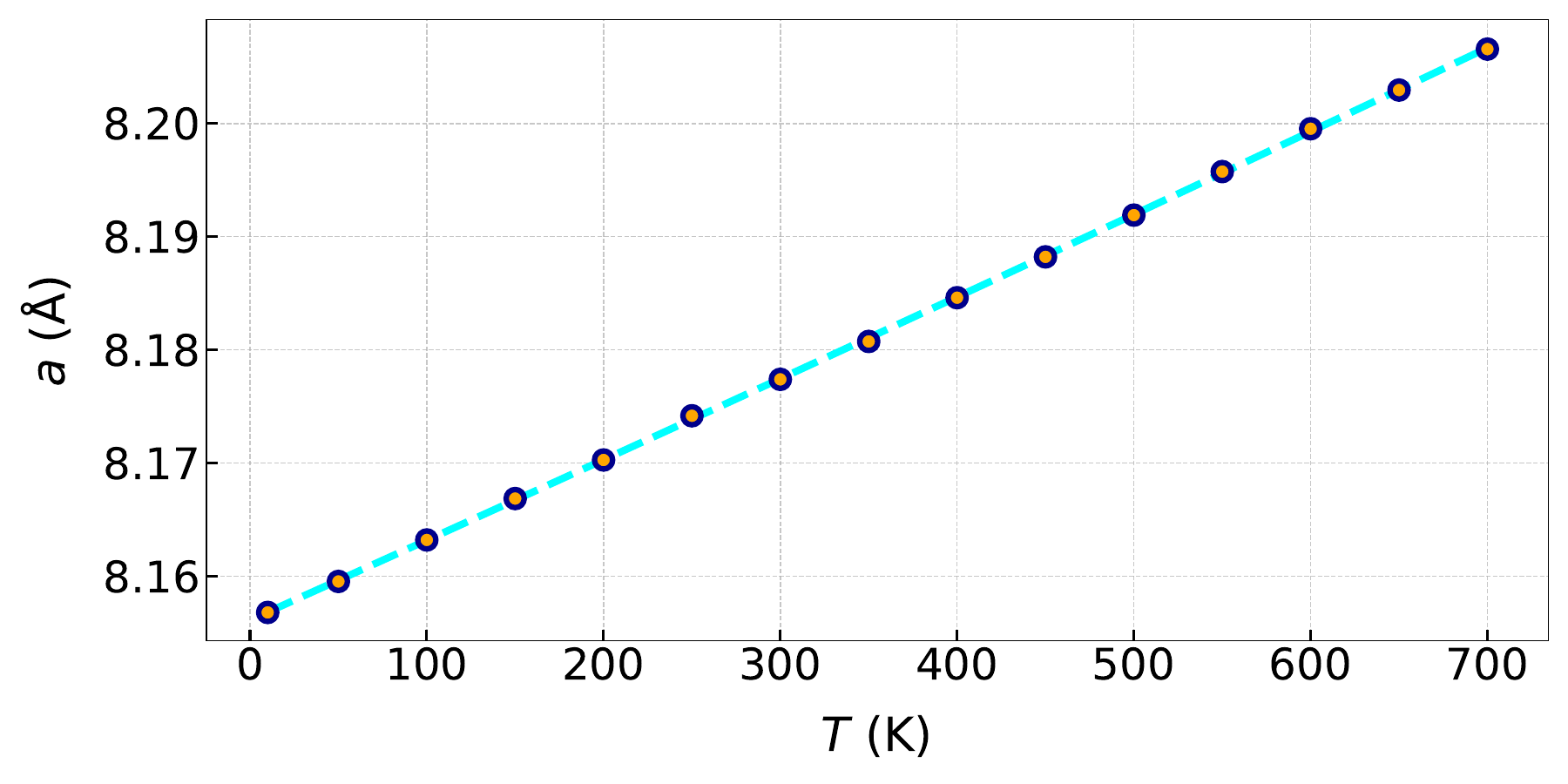}
\caption{Average lattice parameter as a function of temperature for bulk Co$_3$O$_4$ spinel, calculated from fifteen $NPT$ simulations conducted at temperatures ranging from 10~K to 700~K in increments of 50~K. The figure shows a nearly linear increase in the lattice parameter with increasing temperature, a characteristic behavior of crystalline solids due to thermal expansion. The linear thermal expansion coefficient \(\alpha\) can be determined from the slope of this curve, yielding a value of \( \alpha = 8.86 \times 10^{-6} \, \text{K}^{-1} \), which aligns well with experimental data\cite{broemme1991correlation}.}
\label{fig:a-T}
\end{figure}

To determine $\alpha$, fifteen independent $NPT$ MD simulations were run at different temperatures from 10~K up to 700~K in increments of 50~K for 1 ns for a single unit cell. As shown in Figure \ref{fig:a-T} there is a nearly linear increase in the lattice parameter with rising temperature. This behavior is typical for crystalline solids, where the lattice expands as the temperature increases due to the growing vibrational amplitudes of the atoms.

Using Eq.~\ref{eq:thermalexpansion}, the thermal expansion coefficient for Co$_3$O$_4$ over the temperature range from 10~K to 700~K is found to be about \( \alpha = 8.86 \times 10^{-6} \, \text{K}^{-1} \). This result is in excellent agreement with the experimental work of Broemme in 1991 \cite{broemme1991correlation}, in which the thermal expansion coefficient of Co$_3$O$_4$ spinel was reported to be about \( 6 - 9 \times 10^{-6} \, \text{K}^{-1} \) for the temperature range of 380~K to 750~K. It should be emphasized that performing such simulations to extract this value reliably with traditional \textit{ab initio} methods would have been practically impossible due to the high computational costs.

\subsubsection{Extended Simulation Time and Length Scales}\label{subsubsec:time-length}

As the final part, we perform simulations on much longer time and length scales. By doing so, we aim to achieve two objectives: first, to demonstrate the stability of the current HDNNP over significantly extended simulation times and size scales; and second, to verify the reliability of the results obtained in the previous section where a single unit cell and 1 ns simulation time were used by assessing how the reported values might be affected by larger simulation sizes and time scales.

\begin{figure}[h]
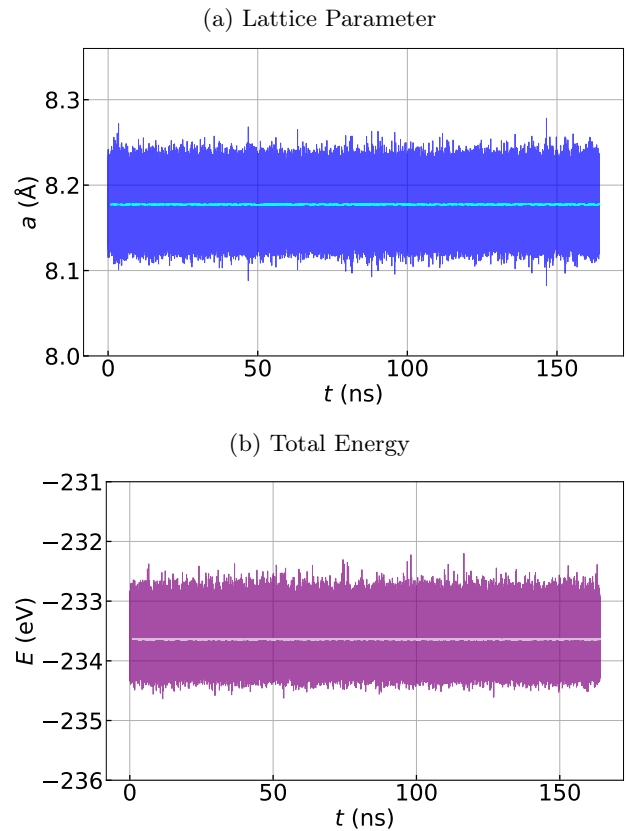

\centering
\begin{subfigure}{0.46\textwidth}
    \centering
    \caption{Lattice Parameter}
    \label{}
    \includegraphics[width=\textwidth, trim=  0 0 0 0, clip=true]{Figures/a_164-ns.png}
\end{subfigure}%
\hfill
\begin{subfigure}{0.46\textwidth}
    \centering
    \caption{Total Energy}
    \label{}
    \includegraphics[width=\textwidth, trim= 0 0 0 0, clip=true]{Figures/E_164-ns.png}
\end{subfigure}
\caption{Evolution of the lattice parameter (a) and total energy (b) of the Co$_3$O$_4$ unit cell over 164 ns of MD simulation in the $NPT$ ensemble at 300~K. The moving averages of both properties are indicated by highlighted lines. The figure demonstrates the stability of both the lattice parameter and total energy throughout the entire simulation duration. The average lattice parameter, 8.177~\AA{}, is nearly identical to the value obtained in the previous section for a 1 ns simulation, with a difference on the order of $6 \times 10^{-5}$~\AA{}, confirming both the long-term stability and the reliability of the earlier results.}
\label{fig:164-ns}
\end{figure}

An MD simulation in the $NPT$ ensemble at 300~K was performed for a bulk Co$_3$O$_4$ unit cell for 164 ns. Figure \ref{fig:164-ns} shows the evolution of the lattice parameter and the total energy of the system over the simulation time. The moving averages values for both the lattice parameter and total energy are shown in the figures with highlighted lines. It can be observed that both properties remain stable throughout the entire 164 ns of simulation. Additionally, the average lattice parameter over 164 ns is 8.177~\AA{}, which is nearly identical to the value reported in the previous section (see Section \ref{subsubsec:Finite-T}) for the 1 ns case at 300~K, with a difference on the order of $6 \times 10^{-5}$~\AA{}. Thus, this confirms both the stability of the simulation with the HDNNP over long time scales and further validates the reliability of the results from the previous section.

Since most simulations so far have been carried out for a (\(1 \times 1 \times 1\)) unit cell containing 56 atoms, we now investigate the convergence of the simulations with respect to system size by repeating the \(NPT\) simulations at temperatures of 10~K, 300~K, and 600~K for a (\(2 \times 2 \times 2\)) supercell containing 448 atoms and a (\(3 \times 3 \times 3\)) supercell containing 1,512 atoms. Each simulation was run for 1~ns. This way we can exclude finite size effects, which may not accurately reflect the true infinite nature of the bulk material. 

\begin{figure}[H]
\centering
\includegraphics[width=0.48\textwidth, trim= 10 10 10 10, clip=true]{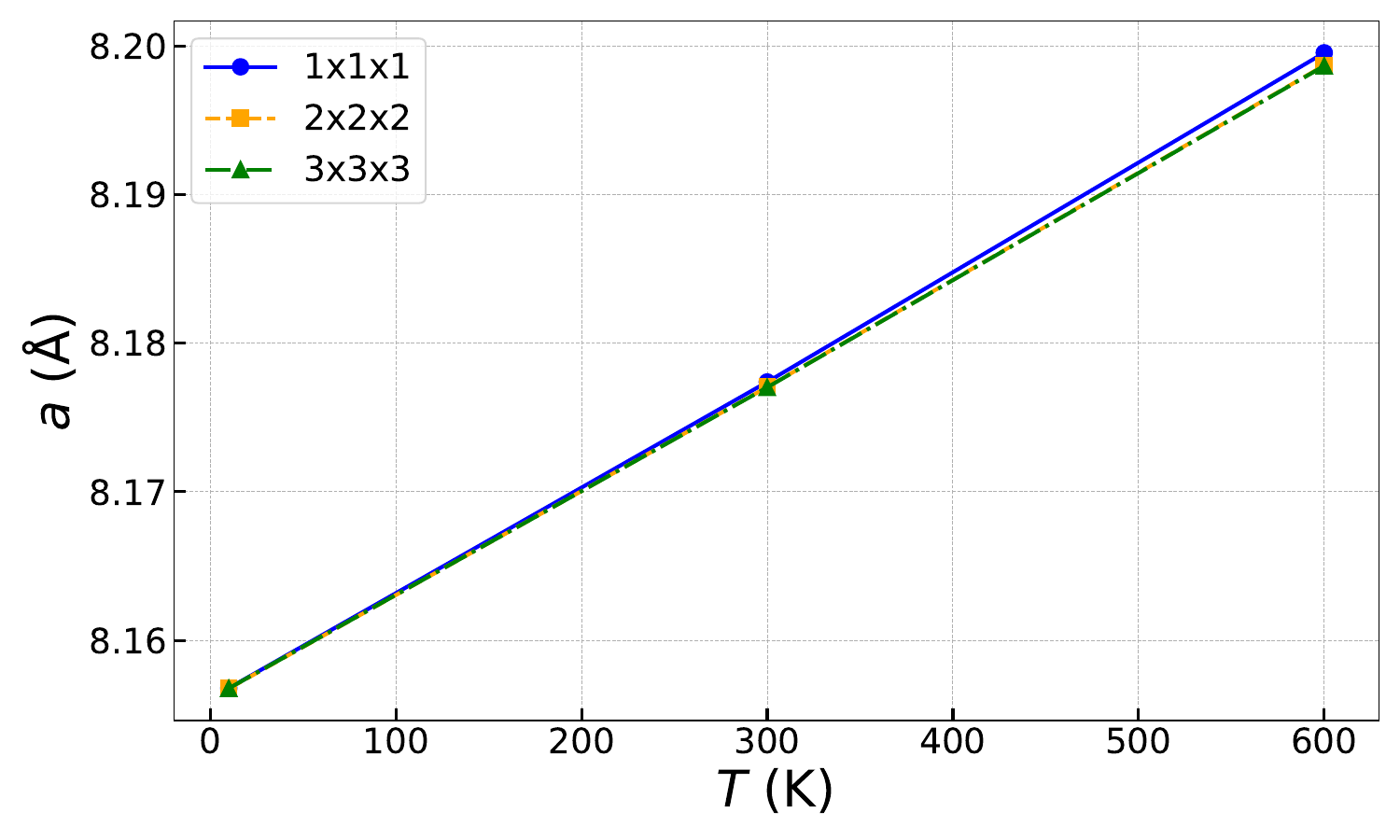}
\caption{Average lattice parameter ($a$) normalized per unit cell over a 1 ns simulation time at 10~K, 300~K, and 600~K for different system sizes of a ($1 \times 1 \times 1$) unit cell (blue), a ($2 \times 2 \times 2$) supercell (orange), and a ($3 \times 3 \times 3$) supercell (green). The results are essentially indistinguishable at 10~K, with slightly larger differences at higher temperatures. The HDNNP predicts normalized lattice parameters of 8.1996~\AA{}, 8.1987~\AA{}, and 8.1987~\AA{} for the respective system sizes at 600~K. The impact of system size on lattice parameters remains minimal, with a negligible effect even at higher temperatures, amounting to only about 0.01\%.}
\label{fig:a-T_effect}
\end{figure}

The resulting average lattice parameters are shown in Figure \ref{fig:a-T_effect}. It can be seen that at 10~K, all three cases predict essentially the same lattice parameters, with differences in the order of $5 \times 10^{-5}$~\AA{}. As the temperature increases, the results slightly differ between the ($1 \times 1 \times 1$) unit cell on one hand, and the ($2 \times 2 \times 2$) and ($3 \times 3 \times 3$) supercells on the other. In fact, normalized per unit cell the HDNNP predicts average values of 8.1996~\AA{}, 8.1987~\AA{}, and 8.1987~\AA{}, respectively, resulting in uncertainties of about 0.0005~\AA{}. It can be concluded that system size differences play a marginal role, and even at elevated temperatures the effect on the lattice constant is only about 0.01 \%. These errors are orders of magnitude lower than the uncertainties obtained with different DFT functionals and how the latter deviate from experimental data (see Section \ref{subsec:lattParam-DFT}). In turn, this underlines that HDNNPs, like MLPs in general, inherit the accuracy from the chosen reference method. So while we have shown that the present HDNNP can reproduce the DFT equilibrium lattice parameter with a very small error, a fundamental limitation remains in that in general GGA functionals tend to overestimate lattice parameters, while the LDA functionals in most cases predict too small values.

In this particular work, optPBE+U overestimates the experimental lattice parameter of 8.087~\AA{} by about 0.85\% (see Section \ref{subsec:lattParam-DFT})). Assuming that vacancies present in real crystals alone cannot be responsible for this remaining error (see Section \ref{subsec:lattParam-DFT})), and that also the thermal expansion plays a minor role, it is likely that the DFT functional contributes most to the discrepancy between theoretical predictions and experimental observations, and the replacement of direct DFT calculations by an MLP contributes only a negligible error. 

The limitations of DFT in predicting material properties have been extensively addressed in the past. To overcome these limitations due to the underlying electronic structure calculations, in the long-term perspective MLPs are very promising as they in principle allow to move towards more accurate electronic structure or post-DFT methods. Methods such as random phase approximation to train the MLP model using delta learning\cite{liu2022phase} as a data-efficient way to achieve this goal, have been recently proposed. In particular, in condensed-phase molecular systems also high-level wave function methods are becoming increasingly more popular~\cite{P6360,P6362}. Realistic finite-temperature simulations enabled by MLPs, such as the one presented in this work for the specific case of Co$_3$O$_4$ spinel, further contribute to more realistic studies of complex systems in chemistry and materials science.

\section{Conclusions}\label{sec:conclusions}

This work presents a high-dimensional neural network potential for bulk Co$_3$O$_4$ spinel. Initially, the electronic structure of Co$_3$O$_4$ was examined, with results showing good agreement with experimental data across various properties when employing the optPBE-vdW functional including a Hubbard U value of 2.43 eV to correct the band gap. Based on this setup, a HDNNP was constructed and validated by
reproducing  basic properties predicted by the underlying optPBE+U functional. It is demonstrated that the HDNNP can accurately provide the lattice parameter, bulk modulus, and phonon properties.

The HDNNP was then used to perform extended molecular dynamics simulations at various temperatures—simulations that would be prohibitively expensive if conducted using traditional \textit{ab initio} molecular dynamics. The effect of temperature on the lattice constant and interatomic distances was particularly explored. It is shown that a 0.023~\AA\ increase in the lattice parameter is expected when going from 300~K to 600~K, closely matching the experimental data. Additionally, the thermal expansion coefficient, $\alpha$, was estimated to be $\alpha = 8.86 \times 10^{-6} \, \text{K}^{-1}$, which is also in excellent agreement with the experiments.

Overall, the HDNNP presented in this work has been shown to be accurate, robust, efficient, and stable for performing a wide range of MD simulations over extended time periods and length scales. It has also been able to calculate various properties without being explicitly fitted to any of them. Additionally, it is demonstrated that the HDNNP's prediction error for the lattice constant is less than 0.01\% compared to the underlying DFT accuracy, while the DFT (optPBE-vdW+U) error in reproducing experimental data is about 0.85\%. This underscores the critical importance of the quality of the underlying electronic structure methods, as machine learning potentials (MLPs) can only be as good as their underlying electronic structure data, inevitably inheriting their limitations.

While this work primarily focused on the bulk properties of the Co$_3$O$_4$ spinel, the presented HDNNP can also serve as a starting point to construct potentials for more complex systems like Co$_3$O$_4$ containing a variety of defects or systems including Co$_3$O$_4$ as a subsystem, such as Co$_3$O$_4$--water interfaces for application in catalysis.

\begin{acknowledgments}
We are grateful for funding by the Deutsche Forschungsgemeinschaft (DFG, German Research Foundation) in TRR/CRC 247 (A10, project-ID 388390466) and under Germany's Excellence Strategy – EXC 2033 RESOLV (project-ID 390677874). The authors also acknowledge the computing time provided to them by the Paderborn Center for Parallel Computing (PC2).
\end{acknowledgments}

\bibliography{literature}

\bibliographystyle{abbrv}

\end{document}


\setstretch{1.0}
\title{Supplementary Information: A High-Dimensional Neural Network Potential for Co$_3$O$_4$}

\author{Amir Omranpour}
\affiliation{Lehrstuhl f\"ur Theoretische Chemie II, Ruhr-Universit\"at Bochum, 44780 Bochum, Germany}
\affiliation{Research Center Chemical Sciences and Sustainability, Research Alliance Ruhr, 44780 Bochum, Germany}
\author{J\"org Behler}
\email{joerg.behler@rub.de}
\affiliation{Lehrstuhl f\"ur Theoretische Chemie II, Ruhr-Universit\"at Bochum, 44780 Bochum, Germany}
\affiliation{Research Center Chemical Sciences and Sustainability, Research Alliance Ruhr, 44780 Bochum, Germany}
\date{\today}

\maketitle

\section{Correlation Plots}\label{sec:correlation}

\begin{figure}[h]
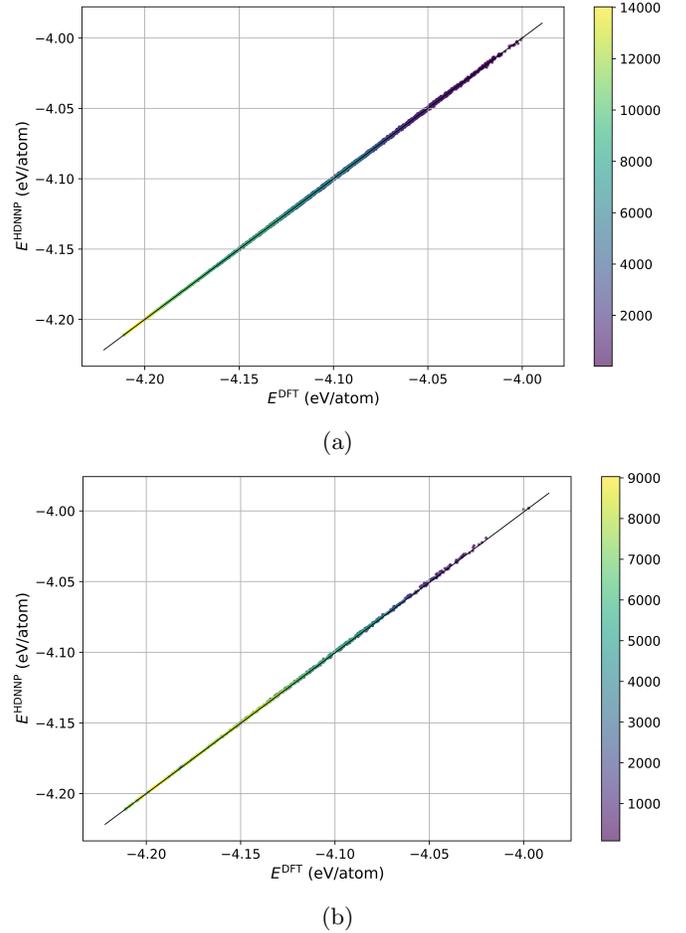

\centering
\begin{subfigure}[b]{0.49\textwidth}
    \centering
    \includegraphics[width=\textwidth, trim= 0 0 0 0, clip=true]{SI-Figures/Energy_train.png}
    \caption{}
    \label{fig:dE_train}
\end{subfigure}
\hfill
\begin{subfigure}[b]{0.49\textwidth}
    \centering
    \includegraphics[width=\textwidth, trim= 0 0 0 0, clip=true]{SI-Figures/Energy_test.png}
    \caption{}
    \label{fig:dE_test}
\end{subfigure}
\caption{Energy correlation plots between the HDNNP predictions and DFT results for (a) the training dataset and (b) the testing dataset. The data points are color-coded based on their relative density, highlighting regions of higher data population.}
\label{fig:dE}
\end{figure}

\noindent Figure \ref{fig:dE} presents the energy correlation plots between the High-Dimensional Neural Network Potential (HDNNP) predictions and the Density Functional Theory (DFT) reference values for both the training and testing datasets. Panel \ref{fig:dE_train} illustrates the energy predictions for the training set, while panel \ref{fig:dE_test} shows the results for the test set. The data points in both plots are color-coded based on their relative density, providing a visual representation of regions with a higher density of data points. The overall alignment of the predicted and reference energy values in these plots demonstrates the ability of the HDNNP model to capture the energy landscape for both training and unseen data without any significant outliers.

\begin{figure}[h]
\centering
\begin{subfigure}[b]{0.49\textwidth}
    \centering
    \includegraphics[width=\textwidth, trim= 0 0 0 0, clip=true]{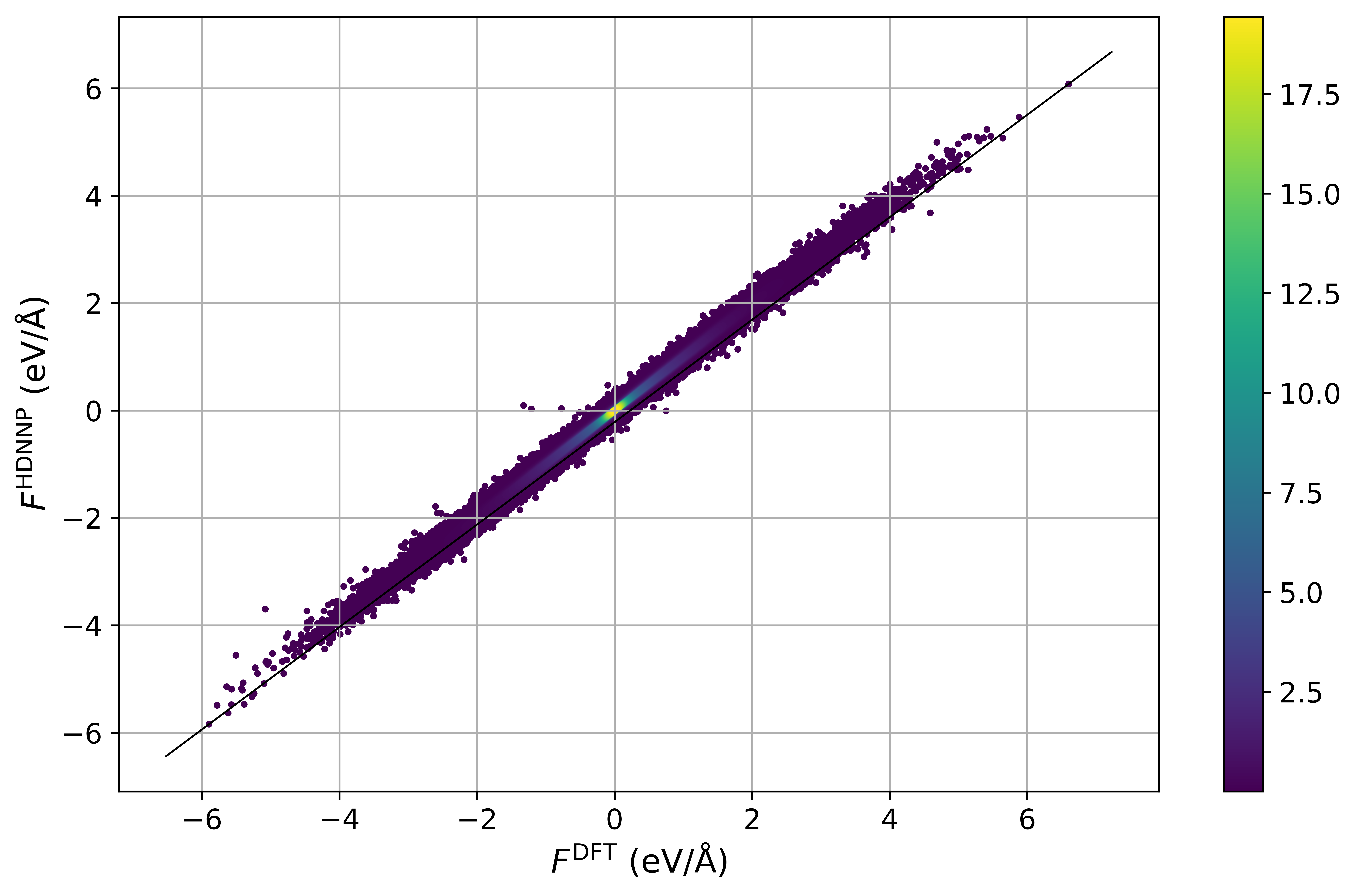}
    \caption{}
    \label{fig:dF_train}
\end{subfigure}
\hfill
\begin{subfigure}[b]{0.49\textwidth}
    \centering
    \includegraphics[width=\textwidth, trim= 0 0 0 0, clip=true]{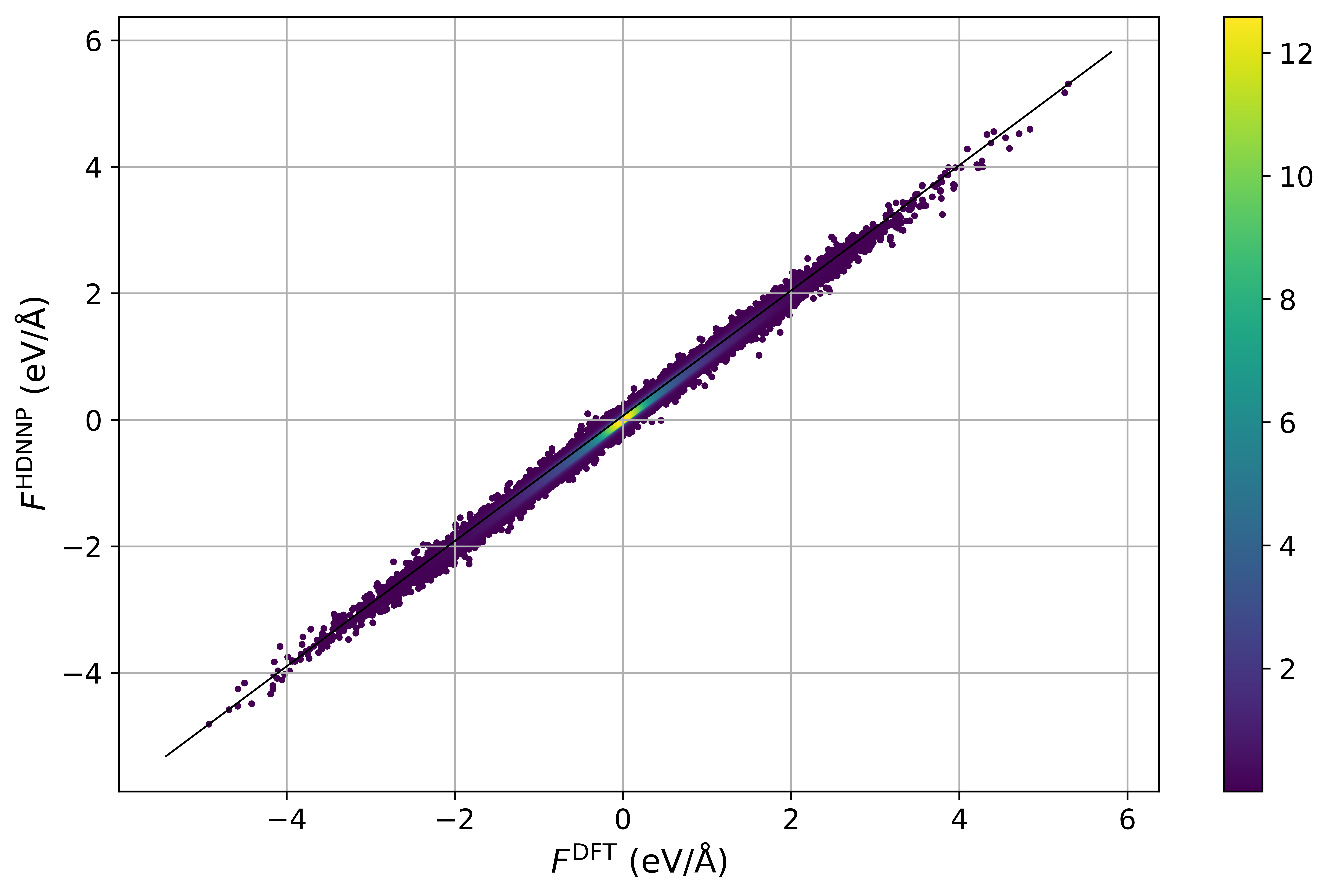}
    \caption{}
    \label{fig:dF_test}
\end{subfigure}
\caption{Force correlation plots between the HDNNP predictions and DFT results for (a) the training dataset and (b) the testing dataset. The data points are color-coded based on their relative density, with brighter regions indicating areas of higher data population.}
\label{fig:dF}
\end{figure}

Similarly, Figure \ref{fig:dF} displays the force correlation plots between the HDNNP predictions and the DFT results for both the training and testing datasets. In panel \ref{fig:dF_train}, the force predictions for the training set are shown, while panel \ref{fig:dF_test} corresponds to the testing set. Again, the data points are color-coded by relative density, with brighter regions signifying areas of higher data point population. These plots allow for a detailed comparison of the predicted forces with the DFT values, highlighting the robustness of the HDNNP model in predicting forces across both datasets. The high degree of correlation in these plots confirms the accuracy of the HDNNP model in replicating the DFT-calculated forces.

\section{Symmetry Functions}

\noindent In this section, the atom-centered symmetry functions (ACSFs) used in the construction of the HDNNP (for modeling the atomic environments of the bulk Co$_3$O$_4$) are presented. These functions are used to capture both the radial and angular aspects of the atomic environment, allowing the network to learn the local environment of each atom.

The radial symmetry functions are characterized by a set of parameters, $\eta$, which control the distance scaling of atomic pairs. Table~\ref{tab:S2} shows the values of $\eta$ employed for different element pairs, e.g., Co--Co, O--O, Co--O, and O--Co interactions. The selected values span a range from 0 to 0.1 a$^{-2}_{0}$, ensuring the ability to model various length scales of atomic interactions within the system.

\begin{table}[h]
\centering
\renewcommand{\arraystretch}{1.5} 
\caption{Parameters $\eta$ of the employed radial atom-centered symmetry functions.}
\label{tab:S2}
\begin{tabular}{cc}
\hline
\textbf{Element pair} & \textbf{$\eta$ /a$^{-2}_{0}$} \\ \hline
Co--Co & 0, 0.0005, 0.005, 0.024, 0.05, 0.1 \\ \hline
O--O   & 0, 0.0005, 0.005, 0.024, 0.05, 0.1 \\ \hline
Co--O  & 0, 0.0005, 0.007, 0.019, 0.05, 0.1 \\ \hline
O--Co  & 0, 0.0005, 0.007, 0.019, 0.05, 0.1 \\ \hline
\end{tabular}
\end{table}

Furthermore, angular symmetry functions capture the relative orientation of triplets of atoms. The angular ACSFs are defined by three key parameters: $\eta$, $\lambda$, and $\zeta$. Table~\ref{tab:S3} provides a list of the angular ACSFs used during the training process, showing the combinations of these parameters for different element triples, e.g., Co--Co--Co, Co--O--O, and others. These symmetry functions, with their respective parameters, form the foundation for accurately representing the local atomic environment in the HDNNP construction for the Co$_3$O$_4$ spinel.

\begin{table}[h]
\centering
\renewcommand{\arraystretch}{1.5}
\caption{List of angular ACSFs, which have been used in the training process starting from a pool of all combinations of the parameters $\lambda = \{-1,1\}$, $\zeta = \{1,2,4,16\}$, and $\eta = \{0,0.025\}$ a$^{-2}_{0}$.}
\label{tab:S3}
\begin{tabular}{cccc}
\hline
\textbf{Element triple} & \textbf{$\eta$ /a$^{-2}_{0}$} & \textbf{$\lambda$} & \textbf{$\zeta$} \\ \hline
Co--Co--Co  & 0, 0.025 &  1, -1 & 1, 2, 4, 16 \\ \hline
Co--O--O    & 0, 0.025 &  1, -1 & 1, 2, 4, 16 \\ \hline
Co--Co--O   & 0, 0.025 &  1, -1 & 1, 2, 4, 16 \\ \hline
O--Co--Co   & 0, 0.025 &  1, -1 & 1, 2, 4, 16 \\ \hline
O--O--Co    & 0, 0.025 &  1, -1 & 1, 2, 4, 16 \\ \hline
O--O--O     & 0, 0.025 &  1, -1 & 1, 2, 4, 16 \\ \hline
\end{tabular}
\end{table}

\section{RuNNer Settings}

\noindent Table~\ref{tab:S4} presents the key settings used in the RuNNer input file for constructing the high-dimensional neural network potential for the Co$_3$O$_4$.

\begin{table}[H]
\centering
\renewcommand{\arraystretch}{1.3}
\caption{Settings in the RuNNer input file for constructing the HDNNP (specification of the ACSFs are left out).}
\label{tab:S4}
\begin{tabular}{ll}
\hline
\textbf{Setting} & \textbf{Value} \\ \hline
nn\_type\_short & 1 \\ \hline
random\_number\_type & 5 \\ \hline
random\_seed & 2000000000 \\ \hline
number\_of\_elements & 2 \\ \hline
elements & Co O \\ \hline
cutoff\_type & 1 \\ \hline
use\_short\_nn & \\ \hline
global\_hidden\_layers\_short & 3 \\ \hline
global\_nodes\_short & 25 20 15 \\ \hline
global\_activation\_short & t t t l \\ \hline
test\_fraction & 0.1 \\ \hline
epochs & 30 \\ \hline
points\_in\_memory & 1000 \\ \hline
mix\_all\_points & \\ \hline
scale\_symmetry\_functions & \\ \hline
center\_symmetry\_functions & \\ \hline
fitting\_unit & eV \\ \hline
precondition\_weights & \\ \hline
use\_short\_forces & \\ \hline
optmode\_short\_energy & 1 \\ \hline
optmode\_short\_force & 1 \\ \hline
kalman\_lambda\_short & 0.98 \\ \hline
kalman\_nue\_short & 0.9987 \\ \hline
short\_energy\_fraction & 1.0 \\ \hline
short\_force\_fraction & 0.1 \\ \hline
weights\_min & -1.0 \\ \hline
weights\_max & 1.0 \\ \hline
nguyen\_widrow\_weights\_short & \\ \hline
short\_force\_error\_threshold & 1.0 \\ \hline
\end{tabular}
\end{table}
